\pdfoutput=1
\documentclass[prapplied,aps,superscriptaddress,showpacs,reprint,longbibliography,nofootinbib]{revtex4-1}
\usepackage{lmodern}
\usepackage[T1]{fontenc}
\usepackage{amsmath,amssymb,amsfonts}
\usepackage{color}
\usepackage{graphicx}
\usepackage[normalem]{ulem}
\usepackage{hyperref}
\usepackage[utf8]{inputenc}
\usepackage{xcolor}
\usepackage{placeins}

\hypersetup{
colorlinks=true,final=true,
        linkcolor=blue,
        citecolor=blue,
        filecolor=blue,
        filecolor=blue,
        urlcolor=blue,
        pdfauthor={S. Krinner and P. Kurpiers},
        pdftitle={Demonstration of an all-microwave controlled-phase gate between far detuned qubits}
}

\begin{document}

\title{Benchmarking Coherent Errors in Controlled-Phase Gates due to Spectator Qubits}


%

\author{S.~Krinner}
\email[]{skrinner@phys.ethz.ch}
\affiliation{Department of Physics, ETH Zurich, 8093 Zurich, Switzerland}

\author{S.~Lazar}
\affiliation{Department of Physics, ETH Zurich, 8093 Zurich, Switzerland}

\author{A.~Remm}
\affiliation{Department of Physics, ETH Zurich, 8093 Zurich, Switzerland}
\author{C.~K.~Andersen}
\affiliation{Department of Physics, ETH Zurich, 8093 Zurich, Switzerland}
\author{N.~Lacroix}
\affiliation{Department of Physics, ETH Zurich, 8093 Zurich, Switzerland}
\author{G.~J.~Norris}
\affiliation{Department of Physics, ETH Zurich, 8093 Zurich, Switzerland}
\author{C.~Hellings}
\affiliation{Department of Physics, ETH Zurich, 8093 Zurich, Switzerland}
\author{M. Gabureac}
\affiliation{Department of Physics, ETH Zurich, 8093 Zurich, Switzerland}
\author{C.~Eichler}
\affiliation{Department of Physics, ETH Zurich, 8093 Zurich, Switzerland}
\author{A.~Wallraff}
\email[]{andreas.wallraff@phys.ethz.ch}
\affiliation{Department of Physics, ETH Zurich, 8093 Zurich, Switzerland}

\date{\today}

\begin{abstract}
A major challenge in operating multi-qubit quantum processors is to mitigate multi-qubit coherent errors. For superconducting circuits, besides crosstalk originating from imperfect isolation of control lines, dispersive coupling between qubits is a major source of multi-qubit coherent errors. We benchmark phase errors in a controlled-phase gate due to dispersive coupling of either of the qubits involved in the gate to one or more spectator qubits. We measure the associated gate infidelity using quantum process tomography. In addition, we point out that, due to coupling of the gate qubits to a non-computational state during the gate, two-qubit conditional phase errors are enhanced. Our work is important for understanding limits to the fidelity of two-qubit gates with finite on/off ratio in multi-qubit settings.

\end{abstract}

\maketitle

\section{Introduction}
In the past two decades the essential building blocks of quantum computers based on superconducting circuits --- high fidelity single- and two-qubit gates, high-fidelity readout and state initialization --- have been developed and steadily improved \cite{Kjaergaard2020a}.
An essential requirement for scaling up present quantum processors towards functional universal quantum computers is to ensure that the performance of individual building blocks is maintained when combining many blocks into a larger processor running operations in parallel. Two-qubit gates are of particular importance since they limit the performance of state-of-the-art quantum processors \cite{Kjaergaard2020a, Gambetta2017}. Although two-qubit gate errors at the $10^{-3}$ level have been demonstrated on few-qubit devices or on isolated parts of multi-qubit devices \cite{Barends2014, Sheldon2016, Rol2019, Barends2019, Foxen2020}, the gate performance typically degrades when operating multiple qubits in parallel for performing larger computations \cite{Gambetta2012, McKay2019, Arute2019, Rudinger2019, Sarovar2019, McKay2020, Wright2019, Erhard2019}. Similar observations are made in quantum processors based on trapped ions \cite{Wright2019, Erhard2019, Gaebler2016, Ballance2016}.

For superconducting circuits, two common reasons for this discrepancy are physical crosstalk originating from imperfect isolation of control lines, and the difficulty of suppressing unwanted couplings between qubits. The latter contains couplings due to spurious electro-magnetic modes as well as couplings present due to finite on/off ratios of two-qubit gates. While isolation of control lines and suppression of spurious electro-magnetic modes can in principle be addressed with careful microwave engineering, finite off-couplings in the form of dispersive couplings \cite{Blais2004,Wallraff2004,Schuster2007a} are characteristic for many of present two-qubit gates \cite{DiCarlo2009,Barends2014,Sheldon2016,McKay2016,Caldwell2018,Rol2019,Barends2019}. While dispersive coupling is key to quantum non-demolition measurements across many physical platforms \cite{Wallraff2005, Thompson2008, Meineke2012, Astner2018, Scarlino2019, Zheng2019},
dispersive coupling in the context of two-qubit gates can lead to coherent errors as well as correlated errors. Both types of errors are known to be particularly harmful in the context of quantum error correction \cite{Gutierrez2016,Greenbaum2018,Bravyi2018,Beale2018,Baireuther2018,Maskara2019}. It is therefore important to characterize those errors to their full extent.

Approaches to reduce dispersive couplings include optimizing gate parameters, such as increasing the frequency detuning between qubits in the idle state, applying dynamical decoupling techniques \cite{Viola1998,Vandersypen2004,Bylander2011,Guo2018a}, and designing more complex passive \cite{McKay2015} or tunable qubit-qubit coupling circuits \cite{Chen2014m,Yan2018b,Mundada2019,Li2019t}. While the dispersive coupling can in principle be brought to zero using tunable coupling circuits with qubits in a certain frequency detuning regime, the overhead in circuit complexity and control hardware is significant. This motivates work to better understand the limitations imposed by dispersive coupling on conventional gate schemes. So far, phase errors due to dispersive coupling have been characterized and mitigated for the constituent qubits in the computational basis \cite{Viola1998,Vandersypen2004,Bylander2011,Guo2018a,Steffen2013,Takita2016,Andersen2019,Bultink2020}. Here, we benchmark errors in the two-qubit conditional phase acquired during a controlled-phase gate due to the dispersive coupling to up to three spectator qubits and measure the associated gate infidelity using quantum process tomography. We show that for understanding the conditional phase error it is necessary to take into account the dispersive shift of the non-computational state involved in the gate.

\section{Dispersive coupling between gate qubits and spectator qubits}
Two-qubit gates are frequently realized by resonantly coupling computational states with each other or with states outside of the computational subspace.
One of the most frequently used two-qubit gates is the family of dynamical flux gates, which includes the resonant iSWAP gate \cite{Bialczak2010,Dewes2012} and the higher-level induced resonant, non-adiabatic \cite{Strauch2003, DiCarlo2010, Barends2019} and adiabatic \cite{DiCarlo2009, Barends2014} controlled-phase gates. They rely on the dynamical flux-tunability of qubit frequencies and are activated by tuning the two-qubit states $|01\rangle$ and $|10\rangle$ into resonance or tuning the $|11\rangle$ and $|02\rangle$ states into resonance, respectively. Here, $|0\rangle, |1\rangle, |2\rangle$ denote the ground, first and second excited states of a transmon qubit. In the idling state, the detuning between the two qubits is much larger than the coupling strength between them, suppressing the resonant interaction. However, a dispersive coupling remains. 
Therefore, any qubit with a physical coupling to the qubits interacting in the gate acts as a spectator qubit, modifying the resonance condition of the gate and thereby inducing gate errors.

The dispersive coupling between two transmon qubits, taken here to be a gate qubit $G$ participating in a two-qubit gate and a spectator qubit $S$, is described by the Hamiltonian
\begin{equation}\label{eqn:Hdisp}
H_{\rm disp}/\hbar = \left(\zeta_{1}\,|1\rangle_G \langle 1|_G + \zeta_{2}\,|2\rangle_G \langle 2|_G \right)|1\rangle_S\langle 1|_S,
\end{equation}
see Appendix \ref{app:Hamiltonian}. The dispersive coupling strengths $\zeta_{1}, \zeta_{2}$ are given by
\begin{align}
\zeta_{1} &=  2 J^2\left(\frac{1}{\Delta+\alpha_S}-\frac{1}{\Delta-\alpha_{G}}\right), \nonumber \\
\zeta_{2} &=  J^2\left(-\frac{1}{\Delta} + \frac{2}{\Delta-\alpha_G} + \frac{3}{\Delta-2\alpha_G} - \frac{4}{\Delta-\alpha_G+\alpha_{S}}\right) \label{eqn:dispShifts}
\end{align}
with the coupling strength $J$, the detuning $\Delta = \omega_S-\omega_G$ between the qubits, and the anharmonicity $\alpha_{G(S)}=(E_{12,G(S)}-E_{01,G(S)})/\hbar$ of the gate (spectator) qubit, with $E_{ij}$ denoting the energy difference between the transmon states $|i\rangle$ and $|j\rangle$. The term with prefactor $\zeta_{1}$ ($\zeta_{2}$) in Eq.~(\ref{eqn:Hdisp}) describes the energy shift of the $|1\rangle$ ($|2\rangle$) state of qubit $G$ conditioned on the state of qubit $S$.

\begin{figure}
     \center
     \includegraphics{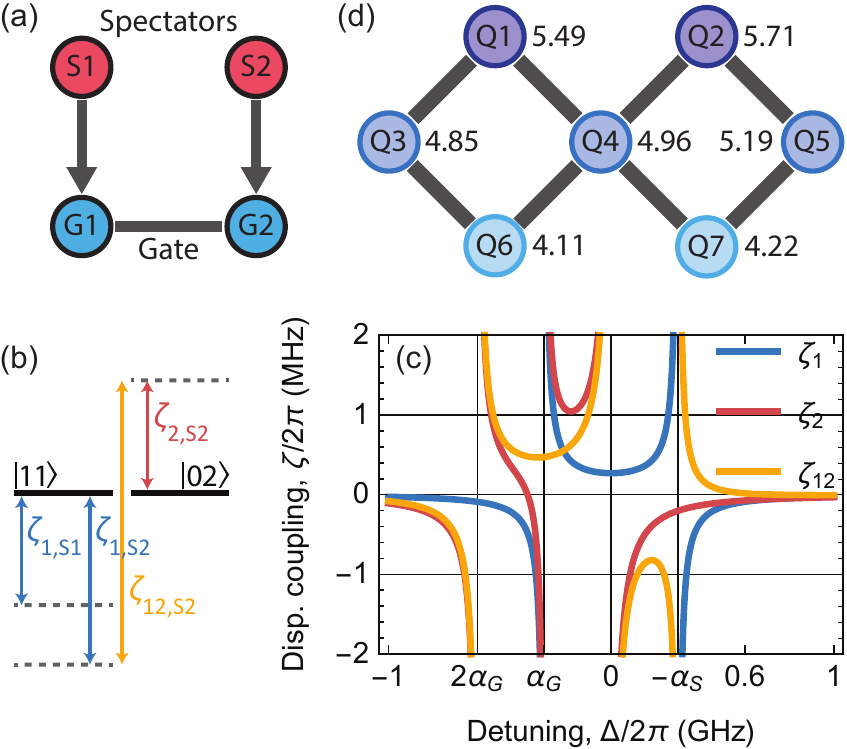}
     \caption{
      (a) Spectator qubits $S1$, $S2$ are coupled to gate qubits $G1$, $G2$ between which we perform a controlled-phase gate. (b) Energy level diagram of the states $|G1G2\rangle=|11\rangle$ and $|02\rangle$, which are shifted due to dispersive interaction with the spectator qubits. (c) Dispersive coupling strengths $\zeta_1$, $\zeta_2$, $\zeta_{12}$ as a function of detuning $\Delta$ between the gate qubits. (d) Qubit connectivity of the studied device. Numbers next to the qubits indicate qubit idling frequencies in GHz.}
     \label{fig1}
\end{figure}

At the heart of the controlled-phase gate is the conditional phase $\Phi_c$ acquired by the $|11\rangle$ state. In the presence of a finite detuning $\delta=(E_{|11\rangle}-E_{|02\rangle})/\hbar$ between the $|11\rangle$ and the $|02\rangle$ states during the gate, $\Phi_c$ deviates from its ideal value of $\pi$. For the non-adiabatic variant of the gate, which has a gate duration $t_g=2\pi/(2\sqrt{2} J)$, it is given by
\begin{equation}\label{eqn:phiC}
\Phi_c = \pi\left(1 + \frac{\delta}{2\sqrt{2}J}\right).
\end{equation}
Hence, a detuning $\delta$ arising from dispersive energy shifts of the $|11\rangle$ and $|02\rangle$ states due to coupling to spectator qubits causes a conditional phase error $\delta\Phi_c=\pi\delta/(2\sqrt{2}J)$.

We consider the generic case where the two qubits $G1$, $G2$ interacting in the controlled-phase gate are coupled to a spectator qubit $S1$ and $S2$, respectively, see Fig.~\ref{fig1}(a). The energy of the state $|11\rangle$ is shifted by the dispersive interaction with the spectator qubits by an amount $\hbar(\zeta_{1,S1} + \zeta_{1,S2})$ with $\zeta_{1,Si}$ denoting the dispersive shift between the $i$th spectator qubit $Si$ and its neighboring gate qubit, see Fig.~\ref{fig1}(b). The energy of the state $|02\rangle$ is only affected by $S2$ and is dispersively shifted by an amount $\hbar\zeta_{2,S2}$. We thus find a dispersive interaction induced detuning of
%
\begin{equation}
	\delta = \zeta_{1,S1} + \zeta_{1,S2} - \zeta_{2,S2}  = \zeta_{1,S1} - \zeta_{12,S2},
\end{equation}
with $\zeta_{12,S2}=\zeta_{2,S2}-\zeta_{1,S2}$ denoting the dispersive shift of the $|1\rangle-|2\rangle$ transition frequency conditioned on the spectator qubit being in the $|1\rangle$ state.
The dispersive couplings $\zeta_{1}$, $\zeta_{2}$ and $\zeta_{12}$ are plotted in Fig.~\ref{fig1}(c) as a function of detuning $\Delta$ for $J/2\pi= 4.5\,$MHz and $\alpha_G=\alpha_S=-300\,$MHz. While $\zeta_1$ has divergences at the two values $\Delta\in \{\alpha_G,\,-\alpha_S$\} due to the Jaynes-Cummings type couplings $|11\rangle\leftrightarrow|20\rangle$ and $|11\rangle\leftrightarrow|02\rangle$, $\zeta_2$ diverges at the four values $\Delta\in \{2\,\alpha_G,\,\alpha_G,\, \alpha_G-\alpha_S,\,0\}$ due to the couplings $|21\rangle\leftrightarrow|30\rangle$, $|20\rangle\leftrightarrow|11\rangle$, $|21\rangle\leftrightarrow|12\rangle$, and $|01\rangle\leftrightarrow|10\rangle$, respectively. All aforementioned resonances must be taken into account for understanding the limitations imposed on the two-qubit gate fidelity by spectator qubits.

\section{Characterization of conditional phase errors}
For our study, we use the seven-qubit device introduced in \cite{Andersen2019b}. The connectivity as well as the idling frequencies of the seven qubits ${\rm Q}i$ have been designed for error detection in the surface code, see Fig.~\ref{fig1}(d) for a schematic. The idling frequencies are chosen to be the sweet spot frequencies, at which the qubits are first-order insensitive to flux noise \cite{Koch2007}. The coupling strength between neighboring qubits is $J/2\pi\approx 4.5(2)\,$MHz. We implement non-adiabatic controlled-phase gates \cite{DiCarlo2010} between any pair of neighbors by applying a unipolar, rectangular current pulse to the flux line of one of the qubits. The flux pulse has a duration $t_g\simeq 80\,$ns and is filtered with Gaussians with $\sigma=1\,$ns.
The anharmonicities of the qubits range from $-290\,$ to $-305$\,MHz.

\begin{figure}[t]
     \center
     \includegraphics{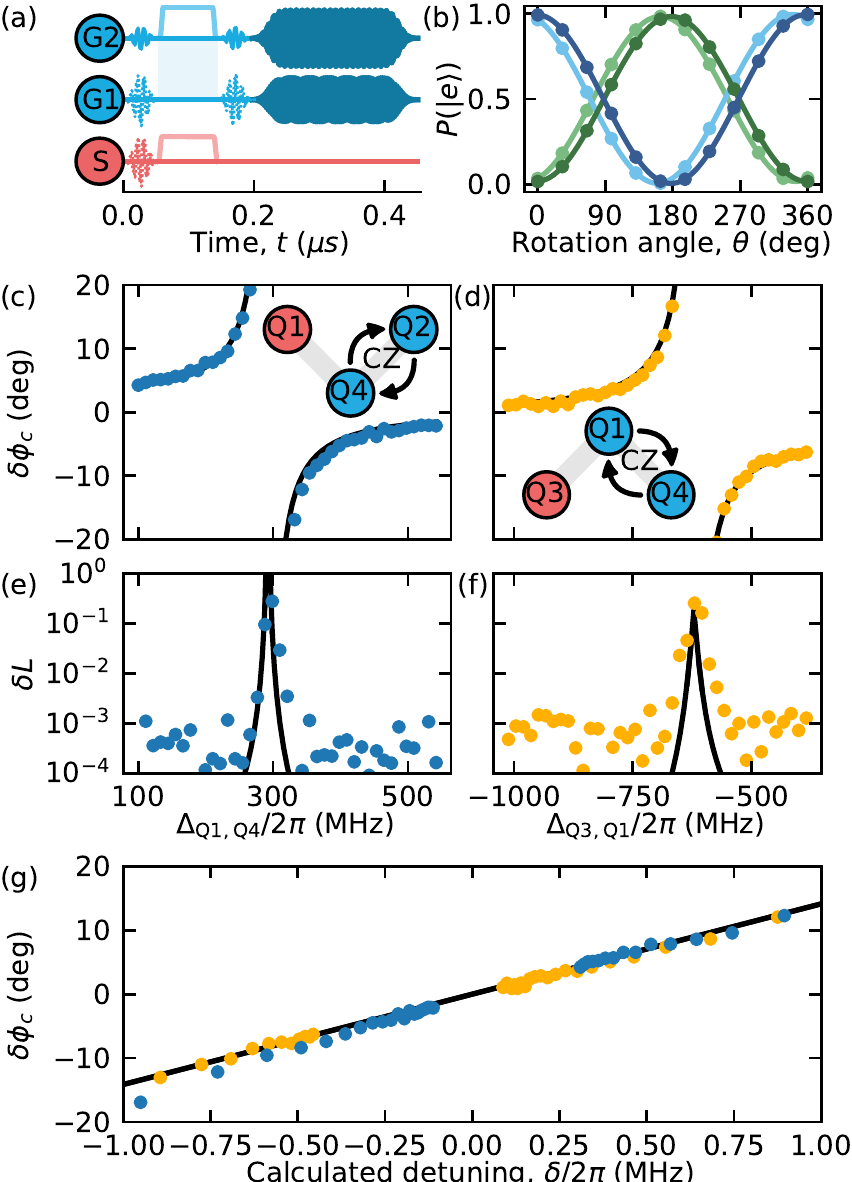}
     \caption{{Conditional phase and leakage errors due to spectator qubits.} (a) Pulse sequence for the conditional phase measurement. Rectangular pulses are flux pulses, short RF bursts represent $\pi/2$ and $\pi$-pulses, long RF bursts at the end represent readout pulses. Dashed pulses indicate that measurements are performed with and without that pulse. (b) Example of a conditional phase measurement with the spectator qubit in $|0\rangle$ (dark blue and dark green data) and in $|1\rangle$ (light blue and light green data). The corresponding sinusoidal fits are shown as solid lines.
     (c) Conditional phase error $\delta \Phi_c$ between ${\rm Q4}$ and ${\rm Q2}$ due to spectator qubit ${\rm Q1}$ and (d) $\delta \Phi_c$ between ${\rm Q1}$ and ${\rm Q4}$ due to spectator qubit ${\rm Q3}$ as a function of the detuning between the spectator qubit and its neighboring gate qubit. (e),(f) Corresponding leakage errors. (g) Conditional phase error $\delta\Phi_c$ from (c) and (d) as a function of detuning between the gate qubit states $|11\rangle$ and $|02\rangle$ during the gate. Solid lines in (c),(d),(g) are calculated based on Eq.~(\ref{eqn:phiC}) and in (e),(f) on Eq.~(\ref{eqn:leakage}).}
     \label{fig2}
\end{figure}

We first study the situation in which the spectator qubit acts on the gate qubit that remains in the computational subspace during the gate. For this purpose, we consider ${\rm Q1}$ as the spectator qubit and ${\rm Q4}$ and ${\rm Q2}$ as the gate qubits, see inset of Fig.~\ref{fig2}(c). We first calibrate the controlled-phase gate with the spectator qubit prepared in $|0\rangle$. 
We measure the conditional phase by performing two Ramsey type experiments on the gate qubit not neighboring the spectator qubit (here $G2={\rm Q2}$), with the other gate qubit (here $G1={\rm Q4}$) prepared in $|0\rangle$ and $|1\rangle$, respectively, see Fig.~\ref{fig2}(a). In each of the experiments the phase of the second $\pi/2$-pulse is varied, resulting in sinusoidal oscillations of the excited state population of ${\rm Q2}$, see Fig.~\ref{fig2}(b). The phase difference between the two oscillations is the conditional phase $\Phi_c$.

We first calibrate amplitude and length of the flux pulse such that $\Phi_c=\pi$. We then repeat the conditional phase measurement with the spectator qubit prepared in $|1\rangle$, and take the difference between the two conditional phase measurements to obtain the conditional phase error $\delta\Phi_{c}$. We average each conditional phase measurement $3.3\times 10^4$ times and interleave in each repetition the measurements with the spectator qubit in $|0\rangle$ and $|1\rangle$ to reduce noise and the susceptibility to parameter drifts. We obtain $\delta\Phi_{c}=-2.1^{\circ}\pm 0.2^{\circ}$, which is in reasonable agreement with the value calculated based on Eq.~(\ref{eqn:phiC}), $\delta\Phi_{c}=\pi\zeta_{1}/(2\sqrt{2}J)=-1.6^{\circ}$.

Next we study the dependence of the conditional phase error on the detuning $\Delta_{\rm Q1, Q4}$ between the spectator qubit ${\rm Q1}$ and the gate qubit ${\rm Q4}$. For this purpose we vary the frequency of the spectator qubit during the conditional phase measurement using a flux pulse applied to the flux line of the spectator qubit. We interleave each conditional phase measurement with a reference measurement in which the spectator qubit is prepared in $|0\rangle$. This is to account for both the spectator state-independent frequency shift of the neighboring gate qubit $Q4$ by $J^2/\Delta_{\rm Q1, Q4}$ and for cross coupling between the flux line of the spectator qubit and the SQUID loops of the gate qubits.
The extracted $\delta\Phi_{c}$ as a function of $\Delta_{\rm Q1, Q4}$ is shown in Fig.~\ref{fig2}(c). We observe that as the detuning is decreased, $\delta\Phi_{c}$ increases and finally diverges at around $\Delta_{\rm Q1, Q4}/2\pi\simeq 289$ MHz, which is the absolute value of the spectator qubit anharmonicity. The data reflects the dependence of $\zeta_{1}$ on $\Delta_{\rm Q1, Q4}$ and is well explained by our model $\delta\Phi_{c}=\pi\zeta_{1}/(2\sqrt{2}J)$ (solid line).

We now turn to the situation where the spectator qubit couples to the gate qubit whose $|2\rangle$ state is involved in the gate. Specifically, we choose ${\rm Q3}$ as the spectator qubit and ${\rm Q1}$ and ${\rm Q4}$ as the gate qubits, see inset of Fig.~\ref{fig2}(d). At the idling frequency of the spectator qubit, corresponding to a detuning $\Delta_{\rm Q3,Q1}/2\pi = -384\,$MHz between spectator and neighboring gate qubit, we measure $\delta\Phi_{c}=-6.3^{\circ}\pm 0.2^{\circ}$, in reasonable agreement with the calculated value $\delta\Phi_{c}=\pi\zeta_{12}/(2\sqrt{2}J)=-6.6^{\circ}$. Analogous to the case described above, we measure the dependence of $\delta\Phi_{c}$ on $\Delta_{\rm Q3,Q1}$, see Fig.~\ref{fig2}(d). $\delta\Phi_c$ increases as we increase $\Delta_{\rm Q3,Q1}$ towards larger negative values, until it diverges and changes sign at $\Delta_{\rm Q3,Q1}/2\pi\simeq -625$\,MHz.
The data is qualitatively described by our model, which shows a resonance at $\Delta_{\rm Q3,Q1}=2\alpha_{\rm Q1}+\beta_{\rm Q1}$ due to resonant coupling of the states $|{\rm Q1Q3}\rangle = |21\rangle$, $|30\rangle$. Here, $\beta_{\rm Q1}=(E_{23}-E_{12})\hbar-(E_{12}-E_{10})/\hbar\approx-35(1)\,$MHz is a correction beyond Eq.~(\ref{eqn:dispShifts}), which takes into account that $E_{23}/\hbar$ differs from $E_{12}/\hbar$ by more than the anharmonicity, see Appendix \ref{app:Hamiltonian}.

Since our model for the conditional phase error, Eq.~(\ref{eqn:phiC}), depends only on the detuning $\delta$ between the states $|11\rangle$ and $|02\rangle$ during the gate, it is instructive to plot $\delta\Phi_c$ as a function of $\delta$ for both acquired data sets, see Fig.~\ref{fig2}(g). Both data sets are well described by the model showing the expected linear dependence of $\delta\Phi_c$ on~$\delta$.

\section{Characterization of leakage errors}
Besides phase errors, a finite detuning $\delta$ during the gate introduces leakage errors, i.e. after the gate a finite fraction $\delta L$ of the population remains in the $|02\rangle$ state. The leakage error for one of the gate qubits prepared in $|0\rangle$+$|1\rangle$ and the other gate qubit prepared in $|1\rangle$ reads
\begin{equation}\label{eqn:leakage}
\delta L = \frac{1}{2}\left(\frac{\pi}{2}\right)^2\left(\frac{\delta}{2\sqrt{2}J}\right)^4.
\end{equation}
The leakage error scales with the fourth power of the small parameter $\delta/J$ and is therefore significantly smaller than the phase errors.

To determine $\delta L$ we measure the $|2\rangle$ state population of $|G2\rangle$ at the end of each conditional phase measurement. Subtracting the value obtained with the spectator qubit prepared in $|0\rangle$ from the value obtained with the spectator qubit in $|1\rangle$ yields $\delta L$. The extracted values of $\delta L$ are shown in Fig.~\ref{fig2}(e) and (f) as a function of the detuning between the spectator qubit and the neighboring gate qubit for the situations corresponding to Fig.~\ref{fig2}(c) and (d). We observe a sizeable leakage error only at detunings corresponding to a divergence of $\delta$, in agreement with a model based on Eq.~(\ref{eqn:leakage}), see solid lines in Fig.~\ref{fig2}(e) and (f). The base line defined by the data corresponds to our measurement accuracy of the $|2\rangle$ state population, which is about $10^{-3}$.

\section{Multiple spectator qubits}
\begin{figure}[t]
     \center
     \includegraphics{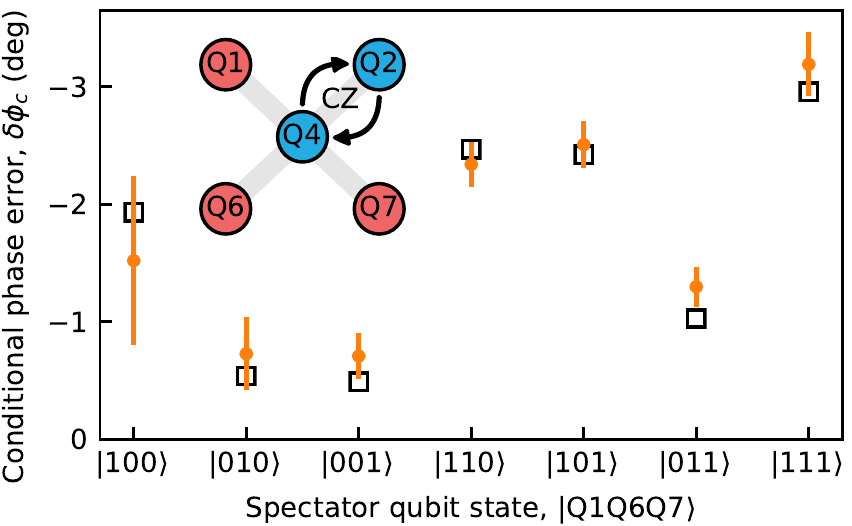}
     \caption{
     Measured conditional phase error $\delta\Phi_c$ (orange circles) in the controlled-phase gate between ${\rm Q4}$ and ${\rm Q2}$ as a function of spectator qubit configuration $|{\rm Q1Q6Q7}\rangle$. Each data point represents the mean of six measurements and error bars indicate one standard deviation. Open black squares are calculated values (see text).}
     \label{fig3}
\end{figure}

Next, we study how errors induced by multiple spectator qubits add up.
We consider the controlled-phase gate between ${\rm Q2}$ and ${\rm Q4}$ and the three spectator qubits ${\rm Q1}$, ${\rm Q6}$, ${\rm Q7}$ coupling to ${\rm Q4}$. After calibrating the gate with all spectator qubits in $|0\rangle$, we measure $\delta\Phi_{c}$ for each of the eight spectator qubit configurations, see orange circles in Fig.~\ref{fig3}. $\delta\Phi_c$ originating from spectator qubit ${\rm Q1}$ is by a factor three larger than $\delta\Phi_c$ originating from ${\rm Q6}$ and ${\rm Q7}$ because the dispersive coupling $\zeta_1$ between ${\rm Q4}$ and ${\rm Q1}$ is larger than between ${\rm Q4}$ and the other two spectator qubits. For the four configurations with multiple spectator qubits in the $|1\rangle$ state, we observe that the measured $\delta\Phi_{c}$ agrees well with the sum over the individual contributions, where only a single spectator qubit is in $|1\rangle$. This shows the coherent nature of the spectator qubit induced conditional phase errors. Overall our measurements agree well with values calculated using Eq.~(\ref{eqn:phiC}) with independently measured values for $\zeta_1$, see black squares in Fig.~\ref{fig3}.

\section{Process tomography}
In addition to the conditional phase error between the gate qubits, the dispersive coupling between a spectator qubit and its neighboring gate qubit also introduces a finite conditional phase between them and therefore mutual dynamical phase errors, see Appendix \ref{app:dynphases} for details and measurements. When considering only the subspace spanned by the gate qubits, this error appears as a single-qubit dynamical phase error $\delta\Phi_d = -\zeta_{1} (t_g+2t_b+t_s)$,
with $t_s=53\,$ns the duration of a single-qubit gate and $t_b=5\,$ns a buffer time which we add before and after the flux pulse inducing the controlled-phase gate.
For the situation corresponding to Fig.~\ref{fig3} we find $\delta\Phi_d\approx -3.5\,\Phi_c$. However, $\delta\Phi_c$ may exceed $\delta\Phi_d$ in absolute value for spectator qubits coupling to the gate qubit whose $|2\rangle$ state is participating in the gate and negative $\Delta_{S,G}$, see also Fig.~\ref{fig1}(c) and Fig.~\ref{fig2}(d).

\begin{figure}[t]
     \center
     \includegraphics{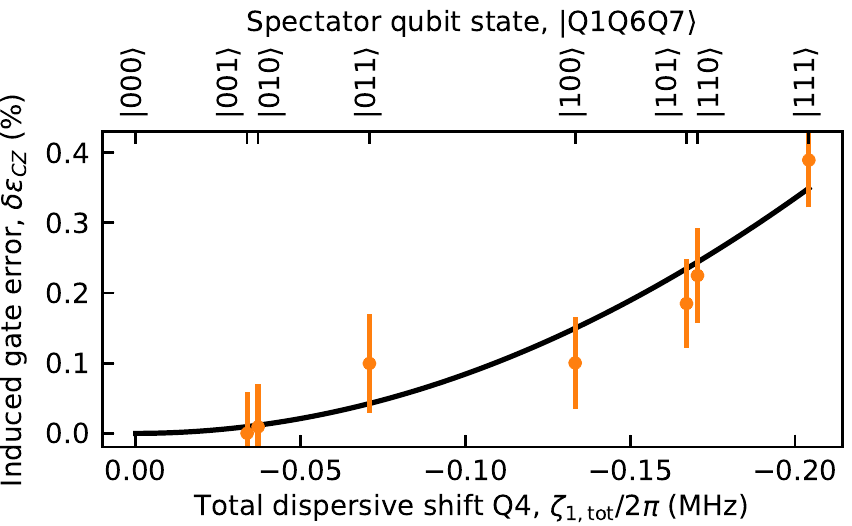}
     \caption{
     Increase in CZ gate error $\delta \varepsilon_{CZ}$ in presence of multiple spectator qubits as a function of the total dispersive shift of the gate qubit ${\rm Q4}$ (orange circles). The top axis indicates the states of the three spectator qubits during each measurement. The solid line is a calculation of the gate error in presence of phase errors only. Error bars are derived from a bootstrapping method.}
     \label{fig4}
\end{figure}

We characterize the joint effect of dynamical and conditional phase errors on the controlled-phase gate between ${\rm Q4}$ and ${\rm Q2}$ by extracting the gate errors $\varepsilon_{CZ}$ from quantum process tomography measurements performed for each of the eight spectator qubit configurations discussed above. By subtracting the gate error from an interleaved reference measurement with all spectator qubits in $|0\rangle$, we obtain the increase in gate error $\delta \varepsilon_{CZ}$. The reference measurements have a mean gate error of $\varepsilon_{CZ}=2.7(2)\,$\%. To increase the signal-to-noise ratio in our measurement of $\delta \varepsilon_{CZ}$ we perform process tomography of three controlled-phase gates executed in series. To obtain $\delta \varepsilon_{CZ}$ of a single controlled-phase gate we divide the obtained gate error increase by nine because the gate error is a quadratic function of the phase errors, see Appendix \ref{app:gateError}.

We show the extracted values of $\delta \varepsilon_{CZ}$ in Fig.~\ref{fig4} as orange points, for each three-spectator-qubit state (top horizontal axis). 
We find that phase errors are responsible for a gate error between 0.0-0.4\% depending on the three-spectator qubit state. The magnitude of the gate errors introduced by spectator qubits is thus comparable to the gate errors of state-of-the-art implementations of two-qubit gates \cite{Barends2014, Sheldon2016, Rol2019, Barends2019, Foxen2020}.

For the case studied here, both the dynamical phase error and the conditional phase error are functions of the total dispersive shift of the gate qubit ${\rm Q4}$. The total dispersive shift of ${\rm Q4}$ is determined by the spectator qubit state $|{\rm Q1Q6Q7}\rangle=|q_1q_6q_7\rangle$ with $q_i\in\{0,1\}$, and reads $\zeta_{1,\rm tot}=q_1\zeta_{1,\rm Q1} + q_6\zeta_{1,\rm Q6} + q_7\zeta_{1,\rm Q7}$. We therefore plot $\delta \varepsilon_{CZ}$ as a function of $\zeta_{1,\rm tot}$, see bottom horizontal axis of Fig.~\ref{fig4}.
We compare our data to a calculation (solid line) of the gate error in presence of coherent phase errors $\delta\Phi_c, \delta\Phi_d$ only, see Appendix~\ref{app:gateError}.
	%
We find that the qualitative dependence of the data $\delta \varepsilon_{CZ}$ on $\zeta_{1, \rm tot}$ is well captured by this model and shows the quadratic increase expected for coherent phase errors.

\section{Conclusion}
To conclude, we have studied how the performance of a controlled-phase gate is affected by the dispersive always-on coupling of the gate qubits to spectator qubits at detunings and coupling strengths typical for our field. We measured conditional phase errors of up to a few degrees, causing gate errors of up to 0.4\,\%. Our results suggest that the widely employed dynamical flux gate needs further conceptual improvement in order to operate at the $10^{-3}-10^{-4}$ error level desired for quantum error correction \cite{Terhal2015n,Reiher2017}. We found that conditional phase errors are particularly pronounced if the spectator qubit has a lower frequency than the gate qubit whose $|2\rangle$ state is involved in the gate. As a remedy, we propose that in such a configuration the detuning between spectator qubit and gate qubit $|\Delta_{S,G}|$ should be chosen to be significantly larger than $|2\alpha_G|$. Finally, we envision that dynamical decoupling of idling spectator qubits can be used to mitigate gate errors.

\FloatBarrier

\section*{Acknowledgments}
The authors acknowledge contributions to the measurement setup from S.~Storz, F.~Swiadek, D. Colao, and T.~Zellweger. The authors acknowledge financial support by the Office of the Director of National Intelligence (ODNI), Intelligence Advanced Research Projects Activity (IARPA), via the U.S. Army Research Office grant W911NF-16-1-0071, by the National Centre of Competence in Research Quantum Science and Technology (NCCR QSIT), a research instrument of the Swiss National Science Foundation (SNSF), by the EU Flagship on Quantum Technology  H2020-FETFLAG-2018-03 project 820363 OpenSuperQ, by the SNFS R'equip grant 206021-170731 and by ETH Zurich. S.~Krinner acknowledges financial support by Fondation Jean-Jacques \& Felicia Lopez-Loreta and the ETH Zurich Foundation. The views and conclusions contained herein are those of the authors and should not be interpreted as necessarily representing the official policies or endorsements, either expressed or implied, of the ODNI, IARPA, or the U.S. Government.

\section*{Author Contributions}
S.K. conceptualized the work. S.K. and S.L. conducted the experiments. S.L. and S.K. analyzed the data. C.K.A. designed the device and S.K., A.R., G.N. and M.G. fabricated the device. C.E. and A.W. supervised the work. S.K., S.L. and A.W. wrote the manuscript with input from all co-authors.


\begin{appendix}

\section{Dispersive Hamiltonian}
\label{app:Hamiltonian}
Any pair of coupled transmon qubits on our seven-qubit device, taken here to be a gate qubit $G$ and a spectator qubit $S$, is described by the Hamiltonian
\begin{align}\label{app:eqn:H}
H/\hbar &= H_0/\hbar + H_I/\hbar \nonumber \\
&= \sum_{i=G,S}\left(\omega_i\,\hat{a}_i^{\dagger}\hat{a}_i + \frac{\alpha_i}{2} \hat{a}_i^{\dagger}\hat{a}_i^{\dagger}\hat{a}_i\hat{a}_i\right) + J(\hat{a}_G\hat{a}_S^{\dagger} + \hat{a}_G^{\dagger}\hat{a}_S) \nonumber \\
\end{align}
with $\hat{a}_{G}$, $\hat{a}_{S}$ ($\hat{a}_{G}^{\dagger}$, $\hat{a}_{S}^{\dagger}$) the lowering (raising) operators of qubits $G$ and $S$, respectively. We diagonalize the Hamiltonian, expand the eigenenergies to second order in $J$, and transform into the rotating frame with respect to $H_0$, i.e.~we subtract the unperturbed eigenenergies of $H_0$ from the diagonal Hamiltonian. The resulting Hamiltonian contains the dispersive interaction terms denoted as $H_{\rm disp}$ in the main text and other dispersive interaction terms not relevant for our study.

To correctly determine the frequency of the resonance of the data shown in Fig.~\ref{fig2}(d) we found it necessary to extend the model Hamiltonian $H_0$, which describes the transmon qubit as an anharmonic oscillator with equally decreasing energy level separation, $E_{i,i+1}=E_{i-1,i}+\alpha$, by the term
\begin{equation}
\sum_{i=G,S}\frac{\beta_i}{6}\hat{a}_i^{\dagger}\hat{a}_i^{\dagger}\hat{a}_i^{\dagger}\hat{a}_i\hat{a}_i\hat{a}_i.
\end{equation}
This term takes into account that the transition frequency from $|2\rangle$ to $|3\rangle$, $E_{23}/\hbar$, differs from $E_{12}/\hbar$ by more than the anharmonicity, i.e. $E_{23}/\hbar=E_{12}/\hbar + \alpha + \beta$. As a consequence, the third term in the equation for $\zeta_2$, Eq.~(\ref{eqn:dispShifts}) of the main text, becomes $3/(\Delta+2\alpha_G+\beta_G)$. We measured $E_{23}/\hbar$ of Q1 using a Ramsey experiment and inferred $\beta_{\rm Q1}=-35(1)$\,MHz, in good agreement with the
calculated value -31.5\,MHz obtained when diagonalizing the transmon Hamiltonian of Q1 \cite{Koch2007}.

\section{Dynamical phase errors}
\label{app:dynphases}
\begin{figure}
     \center
     \includegraphics{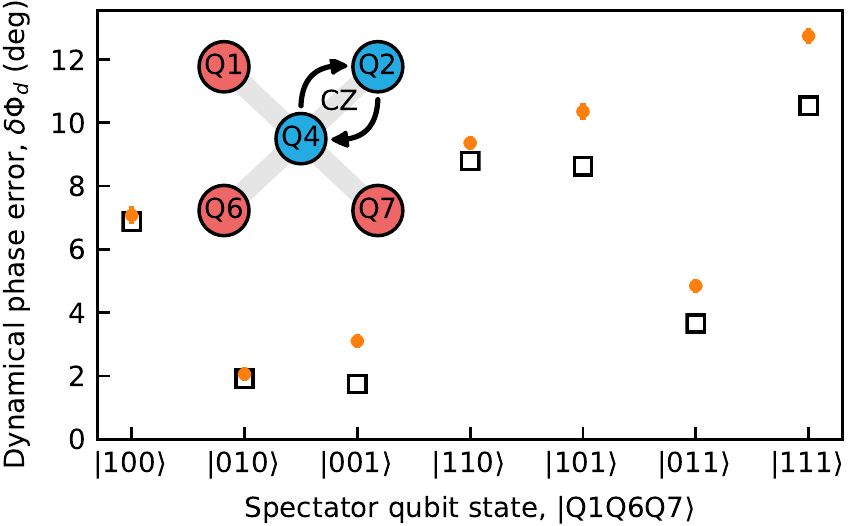}
     \caption{{Dynamical phase errors in presence of multiple spectator qubits.} Dynamical phase error $\delta\Phi_d$ (orange circles) on Q4 during the controlled-phase gate between ${\rm Q4}$ and ${\rm Q2}$ as a function of spectator qubit configuration $|{\rm Q1Q6Q7}\rangle$. Each data point represents the mean of two measurements. Open black squares are calculated values (see text).}
     \label{fig_dynphases}
\end{figure}

We first present measurements of the dynamical single-qubit phase errors occurring on Q4 while it performs a gate with Q2. The origin of the error is a dynamical conditional phase error between Q4 and the spectator qubits Q1, Q6, Q7, which appears as a single-qubit dynamical phase error on Q4 when considering the subspace spanned by the gate qubits Q4 and Q2. For a given spectator qubit state the measurement consists of two Ramsey-type experiments on Q4, one with the given spectator qubit state prepared and one reference experiment with all the spectator qubits prepared in $|0\rangle$. Analogous to the conditional phase measurement, in each experiment we vary the phase of the second $\pi/2$-pulse and extract the accumulated phase by a sinusoidal fit. The phase difference extracted from the two experiments is the dynamical phase error $\delta\Phi_d$. The extracted values are shown in Fig.~\ref{fig_dynphases} for each spectator qubit state. The data agrees well with values calculated from
\begin{equation}
\delta \Phi_{d} = -\zeta_{1, \rm tot} (t_g+2t_b+t_s),
\end{equation}
see black squares. In our calculation, the values of the dispersive shifts $\zeta_{1, \rm Q1}/2\pi=-133(2)\,$kHz, $\zeta_{1, \rm Q6}/2\pi=-37(1)$\,kHz, $\zeta_{1, \rm Q7}/2\pi=-34(1)$\,kHz entering $\zeta_{1, \rm tot}$ are determined from Ramsey experiments.

Since dispersive coupling is mutual, there is also a phase error $\delta\Phi_s$ on the spectator qubits. We measure $\delta\Phi_s$ of Q1 while performing a gate between Q2 and Q4 using a measurement analogous to the one for $\delta\Phi_d$. We obtain $\delta\Phi_s=7.4(5)^{\circ}$ ($\delta\Phi_s=5.8(5)^{\circ}$) with the distant gate qubit Q2 prepared in $|0\rangle$ ($|1\rangle$). The dependence of $\delta\Phi_s$ on the state of the distant gate qubit arises due to the state of the neighboring gate qubit Q4 making a roundtrip between $|1\rangle$ and $|0\rangle$ conditioned on Q2 being in $|1\rangle$. In effect, we expect $\delta\Phi_s=-\zeta_1(t_g+2t_b+t_s)=6.9^{\circ}$ for Q2 prepared in $|0\rangle$ and $\delta\Phi_s=-\zeta_1(0.5t_g+2t_b+t_s)=5.0^{\circ}$ for Q2 prepared in $|1\rangle$, in reasonable agreement with our measurements.

Next, we measure $\delta\Phi_s$ of Q3 while performing a gate between Q1 and Q4. We obtain $\delta\Phi_s=13.1(6)^{\circ}$ ($\delta\Phi_s=7.6(5)^{\circ}$) for Q4 prepared in $|0\rangle$ ($|1\rangle$). Here, the dependence of $\delta\Phi_s$ on the state of the distant gate qubit arises due to the state of the neighboring gate qubit Q1 making a roundtrip between $|1\rangle$ and $|2\rangle$ conditioned on Q4 being in $|1\rangle$. The measured values are in reasonable agreement with the calculated values $\delta\Phi_s=-\zeta_1 t_g-\zeta_{1,\rm id}(2t_b+t_s)=12.9^{\circ}$ and $\delta\Phi_s=-(0.5\zeta_1+0.5\zeta_2)t_g - \zeta_{1,\rm id}(2t_b+t_s)=6.3^{\circ}$, respectively. Here, $\zeta_{1,\rm id}$ denotes the dispersive coupling at the detuning corresponding to the idling frequency of Q1.

\section{Quantum process tomography gate error}
\label{app:gateError}
We calculate the contribution of coherent phase errors to the gate error of a controlled-phase gate. The controlled-phase gate unitary in presence of a conditional phase error $\delta\Phi_c$ and single-qubit dynamic phase errors $\delta\Phi_{d,1}$ and $\delta\Phi_{d,2}$ on the gate qubits $G1$ and $G2$ reads
\begin{align}
	&U_{CZ}(\delta\Phi_{d,1},\delta\Phi_{d,2},\delta\Phi_c) =\\
		&\qquad\begin{pmatrix}
			1 & 0 & 0 & 0\\
			0 & e^{i\delta \Phi_{d,1}} & 0 & 0 \\
			0 & 0 & e^{i\delta \Phi_{d,2}} & 0 \\
			0 & 0 & 0 & e^{i(\pi + \delta \Phi_{c} + \delta \Phi_{d,1} +\delta \Phi_{d,2})}
		\end{pmatrix}, \label{app:eqn:Ucz_th}		
\end{align}
The gate error or infidelity associated with phase errors $\delta\Phi_c$, $\delta\Phi_{d,1}$, $\delta\Phi_{d,2}$ only is given by
\begin{align}
\epsilon_{\rm CZ,P} &= 1-{\rm Trace}[\chi_{CZ}(\delta\Phi_{d,1},\delta\Phi_{d,2},\delta\Phi_c),\chi_{CZ}(0,0,0)] \nonumber \\
&= 0.75 - 0.125\,[{\rm cos}(\delta\Phi_{d,1}) + {\rm cos}(\delta\Phi_{d,2})   \nonumber \\
&\quad + {\rm cos}(\delta\Phi_{d,1}-\delta\Phi_{d,2}) + {\rm cos}(\delta\Phi_{d,1}+\delta\Phi_{d,3})  \nonumber \\
&\quad + {\rm cos}(\delta\Phi_{d,2}+\delta\Phi_{d,3}) + {\rm cos}(\delta\Phi_{d,1}+\delta\Phi_{d,2}+\delta\Phi_{d,3})] \nonumber \\
&\approx 0.25\,\delta\Phi_{d,1}^2 + 0.25\,\delta\Phi_{d,2}^2 + 0.1875\,\delta\Phi_c^2 \nonumber \\
&\quad + 0.25\,\delta\Phi_{d,1}\delta\Phi_c + 0.25\,\delta\Phi_{d,2}\delta\Phi_c, \nonumber
\end{align}
where $\chi_{CZ}(\delta\Phi_{d,1}, \delta\Phi_{d,2}, \delta\Phi_c)$ is the process matrix \cite{Nielsen2011} associated with the two-qubit unitary $U_{CZ}(\delta\Phi_{d,1},\delta\Phi_{d,2},\delta\Phi_c)$. In the last step we have performed a quadratic expansion in $\delta\Phi_{d,1}, \delta\Phi_{d,2}, \delta\Phi_c$. We note that the dynamical and conditional phase errors are each the sum of the individual contributions from each spectator qubit. Furthermore, for the parameter range explored in this study, $\delta\Phi_{d,i}$ and $\delta\Phi_c$ have opposite sign.

For the measurements related to Fig.~\ref{fig4} of the main text we have $\delta\Phi_{d,1}=-\zeta_{1,\rm tot}(t_g+2t_b+t_s)$, $\delta\Phi_{d,2}=0$, $\delta\Phi_{c}=0.5\zeta_{1,\rm tot}t_g$. We find good agreement between our measurements $\delta\epsilon_{\rm CZ}$ and the calculated values $\epsilon_{\rm CZ,P}$, see Fig.~\ref{fig4} of the main text.

\end{appendix}

\bibliography{Q:/USERS/Sebastian//RefDB/QudevRefDB}

\begin{thebibliography}{57}%
\makeatletter
\providecommand \@ifxundefined [1]{%
 \@ifx{#1\undefined}
}%
\providecommand \@ifnum [1]{%
 \ifnum #1\expandafter \@firstoftwo
 \else \expandafter \@secondoftwo
 \fi
}%
\providecommand \@ifx [1]{%
 \ifx #1\expandafter \@firstoftwo
 \else \expandafter \@secondoftwo
 \fi
}%
\providecommand \natexlab [1]{#1}%
\providecommand \enquote  [1]{``#1''}%
\providecommand \bibnamefont  [1]{#1}%
\providecommand \bibfnamefont [1]{#1}%
\providecommand \citenamefont [1]{#1}%
\providecommand \href@noop [0]{\@secondoftwo}%
\providecommand \href [0]{\begingroup \@sanitize@url \@href}%
\providecommand \@href[1]{\@@startlink{#1}\@@href}%
\providecommand \@@href[1]{\endgroup#1\@@endlink}%
\providecommand \@sanitize@url [0]{\catcode `\\12\catcode `\$12\catcode
  `\&12\catcode `\#12\catcode `\^12\catcode `\_12\catcode `\%12\relax}%
\providecommand \@@startlink[1]{}%
\providecommand \@@endlink[0]{}%
\providecommand \url  [0]{\begingroup\@sanitize@url \@url }%
\providecommand \@url [1]{\endgroup\@href {#1}{\urlprefix }}%
\providecommand \urlprefix  [0]{URL }%
\providecommand \Eprint [0]{\href }%
\providecommand \doibase [0]{http://dx.doi.org/}%
\providecommand \selectlanguage [0]{\@gobble}%
\providecommand \bibinfo  [0]{\@secondoftwo}%
\providecommand \bibfield  [0]{\@secondoftwo}%
\providecommand \translation [1]{[#1]}%
\providecommand \BibitemOpen [0]{}%
\providecommand \bibitemStop [0]{}%
\providecommand \bibitemNoStop [0]{.\EOS\space}%
\providecommand \EOS [0]{\spacefactor3000\relax}%
\providecommand \BibitemShut  [1]{\csname bibitem#1\endcsname}%
\let\auto@bib@innerbib\@empty
\bibitem [{\citenamefont {Kjaergaard}\ \emph {et~al.}(2020)\citenamefont
  {Kjaergaard}, \citenamefont {Schwartz}, \citenamefont {Braumüller},
  \citenamefont {Krantz}, \citenamefont {Wang}, \citenamefont {Gustavsson},\
  and\ \citenamefont {Oliver}}]{Kjaergaard2020a}%
  \BibitemOpen
  \bibfield  {author} {\bibinfo {author} {\bibfnamefont {Morten}\ \bibnamefont
  {Kjaergaard}}, \bibinfo {author} {\bibfnamefont {Mollie~E.}\ \bibnamefont
  {Schwartz}}, \bibinfo {author} {\bibfnamefont {Jochen}\ \bibnamefont
  {Braumüller}}, \bibinfo {author} {\bibfnamefont {Philip}\ \bibnamefont
  {Krantz}}, \bibinfo {author} {\bibfnamefont {Joel I.-J.}\ \bibnamefont
  {Wang}}, \bibinfo {author} {\bibfnamefont {Simon}\ \bibnamefont
  {Gustavsson}}, \ and\ \bibinfo {author} {\bibfnamefont {William~D.}\
  \bibnamefont {Oliver}},\ }\bibfield  {title} {\enquote {\bibinfo {title}
  {Superconducting qubits: Current state of play},}\ }\href {\doibase
  10.1146/annurev-conmatphys-031119-050605} {\bibfield  {journal} {\bibinfo
  {journal} {Annual Review of Condensed Matter Physics}\ }\textbf {\bibinfo
  {volume} {11}},\ \bibinfo {pages} {369--395} (\bibinfo {year}
  {2020})}\BibitemShut {NoStop}%
\bibitem [{\citenamefont {Gambetta}\ \emph {et~al.}(2017)\citenamefont
  {Gambetta}, \citenamefont {Chow},\ and\ \citenamefont
  {Steffen}}]{Gambetta2017}%
  \BibitemOpen
  \bibfield  {author} {\bibinfo {author} {\bibfnamefont {Jay~M.}\ \bibnamefont
  {Gambetta}}, \bibinfo {author} {\bibfnamefont {Jerry~M.}\ \bibnamefont
  {Chow}}, \ and\ \bibinfo {author} {\bibfnamefont {Matthias}\ \bibnamefont
  {Steffen}},\ }\bibfield  {title} {\enquote {\bibinfo {title} {Building
  logical qubits in a superconducting quantum computing system},}\ }\href
  {\doibase 10.1038/s41534-016-0004-0} {\bibfield  {journal} {\bibinfo
  {journal} {npj Quantum Information}\ }\textbf {\bibinfo {volume} {3}},\
  \bibinfo {pages} {2} (\bibinfo {year} {2017})}\BibitemShut {NoStop}%
\bibitem [{\citenamefont {Barends}\ \emph {et~al.}(2014)\citenamefont
  {Barends}, \citenamefont {Kelly}, \citenamefont {Megrant}, \citenamefont
  {Veitia}, \citenamefont {Sank}, \citenamefont {Jeffrey}, \citenamefont
  {White}, \citenamefont {Mutus}, \citenamefont {Fowler}, \citenamefont
  {Campbell}, \citenamefont {Chen}, \citenamefont {Chen}, \citenamefont
  {Chiaro}, \citenamefont {Dunsworth}, \citenamefont {Neill}, \citenamefont
  {{O\'Malley}}, \citenamefont {Roushan}, \citenamefont {Vainsencher},
  \citenamefont {Wenner}, \citenamefont {Korotkov}, \citenamefont {Cleland},\
  and\ \citenamefont {Martinis}}]{Barends2014}%
  \BibitemOpen
  \bibfield  {author} {\bibinfo {author} {\bibfnamefont {R.}~\bibnamefont
  {Barends}}, \bibinfo {author} {\bibfnamefont {J.}~\bibnamefont {Kelly}},
  \bibinfo {author} {\bibfnamefont {A.}~\bibnamefont {Megrant}}, \bibinfo
  {author} {\bibfnamefont {A.}~\bibnamefont {Veitia}}, \bibinfo {author}
  {\bibfnamefont {D.}~\bibnamefont {Sank}}, \bibinfo {author} {\bibfnamefont
  {E.}~\bibnamefont {Jeffrey}}, \bibinfo {author} {\bibfnamefont {T.~C.}\
  \bibnamefont {White}}, \bibinfo {author} {\bibfnamefont {J.}~\bibnamefont
  {Mutus}}, \bibinfo {author} {\bibfnamefont {A.~G.}\ \bibnamefont {Fowler}},
  \bibinfo {author} {\bibfnamefont {B.}~\bibnamefont {Campbell}}, \bibinfo
  {author} {\bibfnamefont {Y.}~\bibnamefont {Chen}}, \bibinfo {author}
  {\bibfnamefont {Z.}~\bibnamefont {Chen}}, \bibinfo {author} {\bibfnamefont
  {B.}~\bibnamefont {Chiaro}}, \bibinfo {author} {\bibfnamefont
  {A.}~\bibnamefont {Dunsworth}}, \bibinfo {author} {\bibfnamefont
  {C.}~\bibnamefont {Neill}}, \bibinfo {author} {\bibfnamefont
  {P.}~\bibnamefont {{O\'Malley}}}, \bibinfo {author} {\bibfnamefont
  {P.}~\bibnamefont {Roushan}}, \bibinfo {author} {\bibfnamefont
  {A.}~\bibnamefont {Vainsencher}}, \bibinfo {author} {\bibfnamefont
  {J.}~\bibnamefont {Wenner}}, \bibinfo {author} {\bibfnamefont {A.~N.}\
  \bibnamefont {Korotkov}}, \bibinfo {author} {\bibfnamefont {A.~N.}\
  \bibnamefont {Cleland}}, \ and\ \bibinfo {author} {\bibfnamefont {John~M.}\
  \bibnamefont {Martinis}},\ }\bibfield  {title} {\enquote {\bibinfo {title}
  {Superconducting quantum circuits at the surface code threshold for fault
  tolerance},}\ }\href {\doibase 10.1038/nature13171} {\bibfield  {journal}
  {\bibinfo  {journal} {Nature}\ }\textbf {\bibinfo {volume} {508}},\ \bibinfo
  {pages} {500--503} (\bibinfo {year} {2014})}\BibitemShut {NoStop}%
\bibitem [{\citenamefont {Sheldon}\ \emph {et~al.}(2016)\citenamefont
  {Sheldon}, \citenamefont {Magesan}, \citenamefont {Chow},\ and\ \citenamefont
  {Gambetta}}]{Sheldon2016}%
  \BibitemOpen
  \bibfield  {author} {\bibinfo {author} {\bibfnamefont {Sarah}\ \bibnamefont
  {Sheldon}}, \bibinfo {author} {\bibfnamefont {Easwar}\ \bibnamefont
  {Magesan}}, \bibinfo {author} {\bibfnamefont {Jerry~M.}\ \bibnamefont
  {Chow}}, \ and\ \bibinfo {author} {\bibfnamefont {Jay~M.}\ \bibnamefont
  {Gambetta}},\ }\bibfield  {title} {\enquote {\bibinfo {title} {Procedure for
  systematically tuning up cross-talk in the cross-resonance gate},}\ }\href
  {\doibase 10.1103/PhysRevA.93.060302} {\bibfield  {journal} {\bibinfo
  {journal} {Phys. Rev. A}\ }\textbf {\bibinfo {volume} {93}},\ \bibinfo
  {pages} {060302} (\bibinfo {year} {2016})}\BibitemShut {NoStop}%
\bibitem [{\citenamefont {Rol}\ \emph {et~al.}(2019)\citenamefont {Rol},
  \citenamefont {Battistel}, \citenamefont {Malinowski}, \citenamefont
  {Bultink}, \citenamefont {Tarasinski}, \citenamefont {Vollmer}, \citenamefont
  {Haider}, \citenamefont {Muthusubramanian}, \citenamefont {Bruno},
  \citenamefont {Terhal},\ and\ \citenamefont {DiCarlo}}]{Rol2019}%
  \BibitemOpen
  \bibfield  {author} {\bibinfo {author} {\bibfnamefont {M.~A.}\ \bibnamefont
  {Rol}}, \bibinfo {author} {\bibfnamefont {F.}~\bibnamefont {Battistel}},
  \bibinfo {author} {\bibfnamefont {F.~K.}\ \bibnamefont {Malinowski}},
  \bibinfo {author} {\bibfnamefont {C.~C.}\ \bibnamefont {Bultink}}, \bibinfo
  {author} {\bibfnamefont {B.~M.}\ \bibnamefont {Tarasinski}}, \bibinfo
  {author} {\bibfnamefont {R.}~\bibnamefont {Vollmer}}, \bibinfo {author}
  {\bibfnamefont {N.}~\bibnamefont {Haider}}, \bibinfo {author} {\bibfnamefont
  {N.}~\bibnamefont {Muthusubramanian}}, \bibinfo {author} {\bibfnamefont
  {A.}~\bibnamefont {Bruno}}, \bibinfo {author} {\bibfnamefont {B.~M.}\
  \bibnamefont {Terhal}}, \ and\ \bibinfo {author} {\bibfnamefont
  {L.}~\bibnamefont {DiCarlo}},\ }\bibfield  {title} {\enquote {\bibinfo
  {title} {Fast, high-fidelity conditional-phase gate exploiting leakage
  interference in weakly anharmonic superconducting qubits},}\ }\href {\doibase
  10.1103/PhysRevLett.123.120502} {\bibfield  {journal} {\bibinfo  {journal}
  {Phys. Rev. Lett.}\ }\textbf {\bibinfo {volume} {123}},\ \bibinfo {pages}
  {120502} (\bibinfo {year} {2019})}\BibitemShut {NoStop}%
\bibitem [{\citenamefont {Barends}\ \emph {et~al.}(2019)\citenamefont
  {Barends}, \citenamefont {Quintana}, \citenamefont {Petukhov}, \citenamefont
  {Chen}, \citenamefont {Kafri}, \citenamefont {Kechedzhi}, \citenamefont
  {Collins}, \citenamefont {Naaman}, \citenamefont {Boixo}, \citenamefont
  {Arute}, \citenamefont {Arya}, \citenamefont {Buell}, \citenamefont
  {Burkett}, \citenamefont {Chen}, \citenamefont {Chiaro}, \citenamefont
  {Dunsworth}, \citenamefont {Foxen}, \citenamefont {Fowler}, \citenamefont
  {Gidney}, \citenamefont {Giustina}, \citenamefont {Graff}, \citenamefont
  {Huang}, \citenamefont {Jeffrey}, \citenamefont {Kelly}, \citenamefont
  {Klimov}, \citenamefont {Kostritsa}, \citenamefont {Landhuis}, \citenamefont
  {Lucero}, \citenamefont {McEwen}, \citenamefont {Megrant}, \citenamefont
  {Mi}, \citenamefont {Mutus}, \citenamefont {Neeley}, \citenamefont {Neill},
  \citenamefont {Ostby}, \citenamefont {Roushan}, \citenamefont {Sank},
  \citenamefont {Satzinger}, \citenamefont {Vainsencher}, \citenamefont
  {White}, \citenamefont {Yao}, \citenamefont {Yeh}, \citenamefont {Zalcman},
  \citenamefont {Neven}, \citenamefont {Smelyanskiy},\ and\ \citenamefont
  {Martinis}}]{Barends2019}%
  \BibitemOpen
  \bibfield  {author} {\bibinfo {author} {\bibfnamefont {R.}~\bibnamefont
  {Barends}}, \bibinfo {author} {\bibfnamefont {C.~M.}\ \bibnamefont
  {Quintana}}, \bibinfo {author} {\bibfnamefont {A.~G.}\ \bibnamefont
  {Petukhov}}, \bibinfo {author} {\bibfnamefont {Yu}~\bibnamefont {Chen}},
  \bibinfo {author} {\bibfnamefont {D.}~\bibnamefont {Kafri}}, \bibinfo
  {author} {\bibfnamefont {K.}~\bibnamefont {Kechedzhi}}, \bibinfo {author}
  {\bibfnamefont {R.}~\bibnamefont {Collins}}, \bibinfo {author} {\bibfnamefont
  {O.}~\bibnamefont {Naaman}}, \bibinfo {author} {\bibfnamefont
  {S.}~\bibnamefont {Boixo}}, \bibinfo {author} {\bibfnamefont
  {F.}~\bibnamefont {Arute}}, \bibinfo {author} {\bibfnamefont
  {K.}~\bibnamefont {Arya}}, \bibinfo {author} {\bibfnamefont {D.}~\bibnamefont
  {Buell}}, \bibinfo {author} {\bibfnamefont {B.}~\bibnamefont {Burkett}},
  \bibinfo {author} {\bibfnamefont {Z.}~\bibnamefont {Chen}}, \bibinfo {author}
  {\bibfnamefont {B.}~\bibnamefont {Chiaro}}, \bibinfo {author} {\bibfnamefont
  {A.}~\bibnamefont {Dunsworth}}, \bibinfo {author} {\bibfnamefont
  {B.}~\bibnamefont {Foxen}}, \bibinfo {author} {\bibfnamefont
  {A.}~\bibnamefont {Fowler}}, \bibinfo {author} {\bibfnamefont
  {C.}~\bibnamefont {Gidney}}, \bibinfo {author} {\bibfnamefont
  {M.}~\bibnamefont {Giustina}}, \bibinfo {author} {\bibfnamefont
  {R.}~\bibnamefont {Graff}}, \bibinfo {author} {\bibfnamefont
  {T.}~\bibnamefont {Huang}}, \bibinfo {author} {\bibfnamefont
  {E.}~\bibnamefont {Jeffrey}}, \bibinfo {author} {\bibfnamefont
  {J.}~\bibnamefont {Kelly}}, \bibinfo {author} {\bibfnamefont {P.~V.}\
  \bibnamefont {Klimov}}, \bibinfo {author} {\bibfnamefont {F.}~\bibnamefont
  {Kostritsa}}, \bibinfo {author} {\bibfnamefont {D.}~\bibnamefont {Landhuis}},
  \bibinfo {author} {\bibfnamefont {E.}~\bibnamefont {Lucero}}, \bibinfo
  {author} {\bibfnamefont {M.}~\bibnamefont {McEwen}}, \bibinfo {author}
  {\bibfnamefont {A.}~\bibnamefont {Megrant}}, \bibinfo {author} {\bibfnamefont
  {X.}~\bibnamefont {Mi}}, \bibinfo {author} {\bibfnamefont {J.}~\bibnamefont
  {Mutus}}, \bibinfo {author} {\bibfnamefont {M.}~\bibnamefont {Neeley}},
  \bibinfo {author} {\bibfnamefont {C.}~\bibnamefont {Neill}}, \bibinfo
  {author} {\bibfnamefont {E.}~\bibnamefont {Ostby}}, \bibinfo {author}
  {\bibfnamefont {P.}~\bibnamefont {Roushan}}, \bibinfo {author} {\bibfnamefont
  {D.}~\bibnamefont {Sank}}, \bibinfo {author} {\bibfnamefont {K.~J.}\
  \bibnamefont {Satzinger}}, \bibinfo {author} {\bibfnamefont {A.}~\bibnamefont
  {Vainsencher}}, \bibinfo {author} {\bibfnamefont {T.}~\bibnamefont {White}},
  \bibinfo {author} {\bibfnamefont {J.}~\bibnamefont {Yao}}, \bibinfo {author}
  {\bibfnamefont {P.}~\bibnamefont {Yeh}}, \bibinfo {author} {\bibfnamefont
  {A.}~\bibnamefont {Zalcman}}, \bibinfo {author} {\bibfnamefont
  {H.}~\bibnamefont {Neven}}, \bibinfo {author} {\bibfnamefont {V.~N.}\
  \bibnamefont {Smelyanskiy}}, \ and\ \bibinfo {author} {\bibfnamefont
  {John~M.}\ \bibnamefont {Martinis}},\ }\bibfield  {title} {\enquote {\bibinfo
  {title} {Diabatic gates for frequency-tunable superconducting qubits},}\
  }\href {\doibase 10.1103/PhysRevLett.123.210501} {\bibfield  {journal}
  {\bibinfo  {journal} {Phys. Rev. Lett.}\ }\textbf {\bibinfo {volume} {123}},\
  \bibinfo {pages} {210501} (\bibinfo {year} {2019})}\BibitemShut {NoStop}%
\bibitem [{\citenamefont {Foxen}\ \emph {et~al.}(2020)\citenamefont {Foxen},
  \citenamefont {Neill}, \citenamefont {Dunsworth}, \citenamefont {Roushan},
  \citenamefont {Chiaro}, \citenamefont {Megrant}, \citenamefont {Kelly},
  \citenamefont {Chen}, \citenamefont {Satzinger}, \citenamefont {Barends},
  \citenamefont {Arute}, \citenamefont {Arya}, \citenamefont {Babbush},
  \citenamefont {Bacon}, \citenamefont {Bardin}, \citenamefont {Boixo},
  \citenamefont {Buell}, \citenamefont {Burkett}, \citenamefont {Chen},
  \citenamefont {Collins}, \citenamefont {Farhi}, \citenamefont {Fowler},
  \citenamefont {Gidney}, \citenamefont {Giustina}, \citenamefont {Graff},
  \citenamefont {Harrigan}, \citenamefont {Huang}, \citenamefont {Isakov},
  \citenamefont {Jeffrey}, \citenamefont {Jiang}, \citenamefont {Kafri},
  \citenamefont {Kechedzhi}, \citenamefont {Klimov}, \citenamefont {Korotkov},
  \citenamefont {Kostritsa}, \citenamefont {Landhuis}, \citenamefont {Lucero},
  \citenamefont {McClean}, \citenamefont {McEwen}, \citenamefont {Mi},
  \citenamefont {Mohseni}, \citenamefont {Mutus}, \citenamefont {Naaman},
  \citenamefont {Neeley}, \citenamefont {Niu}, \citenamefont {Petukhov},
  \citenamefont {Quintana}, \citenamefont {Rubin}, \citenamefont {Sank},
  \citenamefont {Smelyanskiy}, \citenamefont {Vainsencher}, \citenamefont
  {White}, \citenamefont {Yao}, \citenamefont {Yeh}, \citenamefont {Zalcman},
  \citenamefont {Neven},\ and\ \citenamefont {Martinis}}]{Foxen2020}%
  \BibitemOpen
  \bibfield  {author} {\bibinfo {author} {\bibfnamefont {B.}~\bibnamefont
  {Foxen}}, \bibinfo {author} {\bibfnamefont {C.}~\bibnamefont {Neill}},
  \bibinfo {author} {\bibfnamefont {A.}~\bibnamefont {Dunsworth}}, \bibinfo
  {author} {\bibfnamefont {P.}~\bibnamefont {Roushan}}, \bibinfo {author}
  {\bibfnamefont {B.}~\bibnamefont {Chiaro}}, \bibinfo {author} {\bibfnamefont
  {A.}~\bibnamefont {Megrant}}, \bibinfo {author} {\bibfnamefont
  {J.}~\bibnamefont {Kelly}}, \bibinfo {author} {\bibfnamefont
  {Z.}~\bibnamefont {Chen}}, \bibinfo {author} {\bibfnamefont {K.}~\bibnamefont
  {Satzinger}}, \bibinfo {author} {\bibfnamefont {R.}~\bibnamefont {Barends}},
  \bibinfo {author} {\bibfnamefont {F.}~\bibnamefont {Arute}}, \bibinfo
  {author} {\bibfnamefont {K.}~\bibnamefont {Arya}}, \bibinfo {author}
  {\bibfnamefont {R.}~\bibnamefont {Babbush}}, \bibinfo {author} {\bibfnamefont
  {D.}~\bibnamefont {Bacon}}, \bibinfo {author} {\bibfnamefont {J.~C.}\
  \bibnamefont {Bardin}}, \bibinfo {author} {\bibfnamefont {S.}~\bibnamefont
  {Boixo}}, \bibinfo {author} {\bibfnamefont {D.}~\bibnamefont {Buell}},
  \bibinfo {author} {\bibfnamefont {B.}~\bibnamefont {Burkett}}, \bibinfo
  {author} {\bibfnamefont {Y.}~\bibnamefont {Chen}}, \bibinfo {author}
  {\bibfnamefont {R.}~\bibnamefont {Collins}}, \bibinfo {author} {\bibfnamefont
  {E.}~\bibnamefont {Farhi}}, \bibinfo {author} {\bibfnamefont
  {A.}~\bibnamefont {Fowler}}, \bibinfo {author} {\bibfnamefont
  {C.}~\bibnamefont {Gidney}}, \bibinfo {author} {\bibfnamefont
  {M.}~\bibnamefont {Giustina}}, \bibinfo {author} {\bibfnamefont
  {R.}~\bibnamefont {Graff}}, \bibinfo {author} {\bibfnamefont
  {M.}~\bibnamefont {Harrigan}}, \bibinfo {author} {\bibfnamefont
  {T.}~\bibnamefont {Huang}}, \bibinfo {author} {\bibfnamefont {S.~V.}\
  \bibnamefont {Isakov}}, \bibinfo {author} {\bibfnamefont {E.}~\bibnamefont
  {Jeffrey}}, \bibinfo {author} {\bibfnamefont {Z.}~\bibnamefont {Jiang}},
  \bibinfo {author} {\bibfnamefont {D.}~\bibnamefont {Kafri}}, \bibinfo
  {author} {\bibfnamefont {K.}~\bibnamefont {Kechedzhi}}, \bibinfo {author}
  {\bibfnamefont {P.}~\bibnamefont {Klimov}}, \bibinfo {author} {\bibfnamefont
  {A.}~\bibnamefont {Korotkov}}, \bibinfo {author} {\bibfnamefont
  {F.}~\bibnamefont {Kostritsa}}, \bibinfo {author} {\bibfnamefont
  {D.}~\bibnamefont {Landhuis}}, \bibinfo {author} {\bibfnamefont
  {E.}~\bibnamefont {Lucero}}, \bibinfo {author} {\bibfnamefont
  {J.}~\bibnamefont {McClean}}, \bibinfo {author} {\bibfnamefont
  {M.}~\bibnamefont {McEwen}}, \bibinfo {author} {\bibfnamefont
  {X.}~\bibnamefont {Mi}}, \bibinfo {author} {\bibfnamefont {M.}~\bibnamefont
  {Mohseni}}, \bibinfo {author} {\bibfnamefont {J.~Y.}\ \bibnamefont {Mutus}},
  \bibinfo {author} {\bibfnamefont {O.}~\bibnamefont {Naaman}}, \bibinfo
  {author} {\bibfnamefont {M.}~\bibnamefont {Neeley}}, \bibinfo {author}
  {\bibfnamefont {M.}~\bibnamefont {Niu}}, \bibinfo {author} {\bibfnamefont
  {A.}~\bibnamefont {Petukhov}}, \bibinfo {author} {\bibfnamefont
  {C.}~\bibnamefont {Quintana}}, \bibinfo {author} {\bibfnamefont
  {N.}~\bibnamefont {Rubin}}, \bibinfo {author} {\bibfnamefont
  {D.}~\bibnamefont {Sank}}, \bibinfo {author} {\bibfnamefont {V.}~\bibnamefont
  {Smelyanskiy}}, \bibinfo {author} {\bibfnamefont {A.}~\bibnamefont
  {Vainsencher}}, \bibinfo {author} {\bibfnamefont {T.~C.}\ \bibnamefont
  {White}}, \bibinfo {author} {\bibfnamefont {Z.}~\bibnamefont {Yao}}, \bibinfo
  {author} {\bibfnamefont {P.}~\bibnamefont {Yeh}}, \bibinfo {author}
  {\bibfnamefont {A.}~\bibnamefont {Zalcman}}, \bibinfo {author} {\bibfnamefont
  {H.}~\bibnamefont {Neven}}, \ and\ \bibinfo {author} {\bibfnamefont {J.~M.}\
  \bibnamefont {Martinis}},\ }\bibfield  {title} {\enquote {\bibinfo {title}
  {Demonstrating a continuous set of two-qubit gates for near-term quantum
  algorithms},}\ }\href {https://arxiv.org/abs/2001.08343} {\bibfield
  {journal} {\bibinfo  {journal} {arXiv:2001.08343}\ } (\bibinfo {year}
  {2020})}\BibitemShut {NoStop}%
\bibitem [{\citenamefont {Gambetta}\ \emph {et~al.}(2012)\citenamefont
  {Gambetta}, \citenamefont {C\'orcoles}, \citenamefont {Merkel}, \citenamefont
  {Johnson}, \citenamefont {Smolin}, \citenamefont {Chow}, \citenamefont
  {Ryan}, \citenamefont {Rigetti}, \citenamefont {Poletto}, \citenamefont
  {Ohki}, \citenamefont {Ketchen},\ and\ \citenamefont
  {Steffen}}]{Gambetta2012}%
  \BibitemOpen
  \bibfield  {author} {\bibinfo {author} {\bibfnamefont {Jay~M.}\ \bibnamefont
  {Gambetta}}, \bibinfo {author} {\bibfnamefont {A.~D.}\ \bibnamefont
  {C\'orcoles}}, \bibinfo {author} {\bibfnamefont {S.~T.}\ \bibnamefont
  {Merkel}}, \bibinfo {author} {\bibfnamefont {B.~R.}\ \bibnamefont {Johnson}},
  \bibinfo {author} {\bibfnamefont {John~A.}\ \bibnamefont {Smolin}}, \bibinfo
  {author} {\bibfnamefont {Jerry~M.}\ \bibnamefont {Chow}}, \bibinfo {author}
  {\bibfnamefont {Colm~A.}\ \bibnamefont {Ryan}}, \bibinfo {author}
  {\bibfnamefont {Chad}\ \bibnamefont {Rigetti}}, \bibinfo {author}
  {\bibfnamefont {S.}~\bibnamefont {Poletto}}, \bibinfo {author} {\bibfnamefont
  {Thomas~A.}\ \bibnamefont {Ohki}}, \bibinfo {author} {\bibfnamefont
  {Mark~B.}\ \bibnamefont {Ketchen}}, \ and\ \bibinfo {author} {\bibfnamefont
  {M.}~\bibnamefont {Steffen}},\ }\bibfield  {title} {\enquote {\bibinfo
  {title} {Characterization of addressability by simultaneous randomized
  benchmarking},}\ }\href {\doibase 10.1103/PhysRevLett.109.240504} {\bibfield
  {journal} {\bibinfo  {journal} {Phys. Rev. Lett.}\ }\textbf {\bibinfo
  {volume} {109}},\ \bibinfo {pages} {240504} (\bibinfo {year}
  {2012})}\BibitemShut {NoStop}%
\bibitem [{\citenamefont {McKay}\ \emph {et~al.}(2019)\citenamefont {McKay},
  \citenamefont {Sheldon}, \citenamefont {Smolin}, \citenamefont {Chow},\ and\
  \citenamefont {Gambetta}}]{McKay2019}%
  \BibitemOpen
  \bibfield  {author} {\bibinfo {author} {\bibfnamefont {David~C.}\
  \bibnamefont {McKay}}, \bibinfo {author} {\bibfnamefont {Sarah}\ \bibnamefont
  {Sheldon}}, \bibinfo {author} {\bibfnamefont {John~A.}\ \bibnamefont
  {Smolin}}, \bibinfo {author} {\bibfnamefont {Jerry~M.}\ \bibnamefont {Chow}},
  \ and\ \bibinfo {author} {\bibfnamefont {Jay~M.}\ \bibnamefont {Gambetta}},\
  }\bibfield  {title} {\enquote {\bibinfo {title} {Three-qubit randomized
  benchmarking},}\ }\href {\doibase 10.1103/PhysRevLett.122.200502} {\bibfield
  {journal} {\bibinfo  {journal} {Phys. Rev. Lett.}\ }\textbf {\bibinfo
  {volume} {122}},\ \bibinfo {pages} {200502} (\bibinfo {year}
  {2019})}\BibitemShut {NoStop}%
\bibitem [{\citenamefont {Arute}\ \emph {et~al.}(2019)\citenamefont {Arute},
  \citenamefont {Arya}, \citenamefont {Babbush}, \citenamefont {Bacon},
  \citenamefont {Bardin}, \citenamefont {Barends}, \citenamefont {Biswas},
  \citenamefont {Boixo}, \citenamefont {Brandao}, \citenamefont {Buell},
  \citenamefont {Burkett}, \citenamefont {Chen}, \citenamefont {Chen},
  \citenamefont {Chiaro}, \citenamefont {Collins}, \citenamefont {Courtney},
  \citenamefont {Dunsworth}, \citenamefont {Farhi}, \citenamefont {Foxen},
  \citenamefont {Fowler}, \citenamefont {Gidney}, \citenamefont {Giustina},
  \citenamefont {Graff}, \citenamefont {Guerin}, \citenamefont {Habegger},
  \citenamefont {Harrigan}, \citenamefont {Hartmann}, \citenamefont {Ho},
  \citenamefont {Hoffmann}, \citenamefont {Huang}, \citenamefont {Humble},
  \citenamefont {Isakov}, \citenamefont {Jeffrey}, \citenamefont {Jiang},
  \citenamefont {Kafri}, \citenamefont {Kechedzhi}, \citenamefont {Kelly},
  \citenamefont {Klimov}, \citenamefont {Knysh}, \citenamefont {Korotkov},
  \citenamefont {Kostritsa}, \citenamefont {Landhuis}, \citenamefont
  {Lindmark}, \citenamefont {Lucero}, \citenamefont {Lyakh}, \citenamefont
  {Mandrà}, \citenamefont {McClean}, \citenamefont {McEwen}, \citenamefont
  {Megrant}, \citenamefont {Mi}, \citenamefont {Michielsen}, \citenamefont
  {Mohseni}, \citenamefont {Mutus}, \citenamefont {Naaman}, \citenamefont
  {Neeley}, \citenamefont {Neill}, \citenamefont {Niu}, \citenamefont {Ostby},
  \citenamefont {Petukhov}, \citenamefont {Platt}, \citenamefont {Quintana},
  \citenamefont {Rieffel}, \citenamefont {Roushan}, \citenamefont {Rubin},
  \citenamefont {Sank}, \citenamefont {Satzinger}, \citenamefont {Smelyanskiy},
  \citenamefont {Sung}, \citenamefont {Trevithick}, \citenamefont
  {Vainsencher}, \citenamefont {Villalonga}, \citenamefont {White},
  \citenamefont {Yao}, \citenamefont {Yeh}, \citenamefont {Zalcman},
  \citenamefont {Neven},\ and\ \citenamefont {Martinis}}]{Arute2019}%
  \BibitemOpen
  \bibfield  {author} {\bibinfo {author} {\bibfnamefont {Frank}\ \bibnamefont
  {Arute}}, \bibinfo {author} {\bibfnamefont {Kunal}\ \bibnamefont {Arya}},
  \bibinfo {author} {\bibfnamefont {Ryan}\ \bibnamefont {Babbush}}, \bibinfo
  {author} {\bibfnamefont {Dave}\ \bibnamefont {Bacon}}, \bibinfo {author}
  {\bibfnamefont {Joseph~C.}\ \bibnamefont {Bardin}}, \bibinfo {author}
  {\bibfnamefont {Rami}\ \bibnamefont {Barends}}, \bibinfo {author}
  {\bibfnamefont {Rupak}\ \bibnamefont {Biswas}}, \bibinfo {author}
  {\bibfnamefont {Sergio}\ \bibnamefont {Boixo}}, \bibinfo {author}
  {\bibfnamefont {Fernando G. S.~L.}\ \bibnamefont {Brandao}}, \bibinfo
  {author} {\bibfnamefont {David~A.}\ \bibnamefont {Buell}}, \bibinfo {author}
  {\bibfnamefont {Brian}\ \bibnamefont {Burkett}}, \bibinfo {author}
  {\bibfnamefont {Yu}~\bibnamefont {Chen}}, \bibinfo {author} {\bibfnamefont
  {Zijun}\ \bibnamefont {Chen}}, \bibinfo {author} {\bibfnamefont {Ben}\
  \bibnamefont {Chiaro}}, \bibinfo {author} {\bibfnamefont {Roberto}\
  \bibnamefont {Collins}}, \bibinfo {author} {\bibfnamefont {William}\
  \bibnamefont {Courtney}}, \bibinfo {author} {\bibfnamefont {Andrew}\
  \bibnamefont {Dunsworth}}, \bibinfo {author} {\bibfnamefont {Edward}\
  \bibnamefont {Farhi}}, \bibinfo {author} {\bibfnamefont {Brooks}\
  \bibnamefont {Foxen}}, \bibinfo {author} {\bibfnamefont {Austin}\
  \bibnamefont {Fowler}}, \bibinfo {author} {\bibfnamefont {Craig}\
  \bibnamefont {Gidney}}, \bibinfo {author} {\bibfnamefont {Marissa}\
  \bibnamefont {Giustina}}, \bibinfo {author} {\bibfnamefont {Rob}\
  \bibnamefont {Graff}}, \bibinfo {author} {\bibfnamefont {Keith}\ \bibnamefont
  {Guerin}}, \bibinfo {author} {\bibfnamefont {Steve}\ \bibnamefont
  {Habegger}}, \bibinfo {author} {\bibfnamefont {Matthew~P.}\ \bibnamefont
  {Harrigan}}, \bibinfo {author} {\bibfnamefont {Michael~J.}\ \bibnamefont
  {Hartmann}}, \bibinfo {author} {\bibfnamefont {Alan}\ \bibnamefont {Ho}},
  \bibinfo {author} {\bibfnamefont {Markus}\ \bibnamefont {Hoffmann}}, \bibinfo
  {author} {\bibfnamefont {Trent}\ \bibnamefont {Huang}}, \bibinfo {author}
  {\bibfnamefont {Travis~S.}\ \bibnamefont {Humble}}, \bibinfo {author}
  {\bibfnamefont {Sergei~V.}\ \bibnamefont {Isakov}}, \bibinfo {author}
  {\bibfnamefont {Evan}\ \bibnamefont {Jeffrey}}, \bibinfo {author}
  {\bibfnamefont {Zhang}\ \bibnamefont {Jiang}}, \bibinfo {author}
  {\bibfnamefont {Dvir}\ \bibnamefont {Kafri}}, \bibinfo {author}
  {\bibfnamefont {Kostyantyn}\ \bibnamefont {Kechedzhi}}, \bibinfo {author}
  {\bibfnamefont {Julian}\ \bibnamefont {Kelly}}, \bibinfo {author}
  {\bibfnamefont {Paul~V.}\ \bibnamefont {Klimov}}, \bibinfo {author}
  {\bibfnamefont {Sergey}\ \bibnamefont {Knysh}}, \bibinfo {author}
  {\bibfnamefont {Alexander}\ \bibnamefont {Korotkov}}, \bibinfo {author}
  {\bibfnamefont {Fedor}\ \bibnamefont {Kostritsa}}, \bibinfo {author}
  {\bibfnamefont {David}\ \bibnamefont {Landhuis}}, \bibinfo {author}
  {\bibfnamefont {Mike}\ \bibnamefont {Lindmark}}, \bibinfo {author}
  {\bibfnamefont {Erik}\ \bibnamefont {Lucero}}, \bibinfo {author}
  {\bibfnamefont {Dmitry}\ \bibnamefont {Lyakh}}, \bibinfo {author}
  {\bibfnamefont {Salvatore}\ \bibnamefont {Mandrà}}, \bibinfo {author}
  {\bibfnamefont {Jarrod~R.}\ \bibnamefont {McClean}}, \bibinfo {author}
  {\bibfnamefont {Matthew}\ \bibnamefont {McEwen}}, \bibinfo {author}
  {\bibfnamefont {Anthony}\ \bibnamefont {Megrant}}, \bibinfo {author}
  {\bibfnamefont {Xiao}\ \bibnamefont {Mi}}, \bibinfo {author} {\bibfnamefont
  {Kristel}\ \bibnamefont {Michielsen}}, \bibinfo {author} {\bibfnamefont
  {Masoud}\ \bibnamefont {Mohseni}}, \bibinfo {author} {\bibfnamefont {Josh}\
  \bibnamefont {Mutus}}, \bibinfo {author} {\bibfnamefont {Ofer}\ \bibnamefont
  {Naaman}}, \bibinfo {author} {\bibfnamefont {Matthew}\ \bibnamefont
  {Neeley}}, \bibinfo {author} {\bibfnamefont {Charles}\ \bibnamefont {Neill}},
  \bibinfo {author} {\bibfnamefont {Murphy~Yuezhen}\ \bibnamefont {Niu}},
  \bibinfo {author} {\bibfnamefont {Eric}\ \bibnamefont {Ostby}}, \bibinfo
  {author} {\bibfnamefont {Andre}\ \bibnamefont {Petukhov}}, \bibinfo {author}
  {\bibfnamefont {John~C.}\ \bibnamefont {Platt}}, \bibinfo {author}
  {\bibfnamefont {Chris}\ \bibnamefont {Quintana}}, \bibinfo {author}
  {\bibfnamefont {Eleanor~G.}\ \bibnamefont {Rieffel}}, \bibinfo {author}
  {\bibfnamefont {Pedram}\ \bibnamefont {Roushan}}, \bibinfo {author}
  {\bibfnamefont {Nicholas~C.}\ \bibnamefont {Rubin}}, \bibinfo {author}
  {\bibfnamefont {Daniel}\ \bibnamefont {Sank}}, \bibinfo {author}
  {\bibfnamefont {Kevin~J.}\ \bibnamefont {Satzinger}}, \bibinfo {author}
  {\bibfnamefont {Vadim}\ \bibnamefont {Smelyanskiy}}, \bibinfo {author}
  {\bibfnamefont {Kevin~J.}\ \bibnamefont {Sung}}, \bibinfo {author}
  {\bibfnamefont {Matthew~D.}\ \bibnamefont {Trevithick}}, \bibinfo {author}
  {\bibfnamefont {Amit}\ \bibnamefont {Vainsencher}}, \bibinfo {author}
  {\bibfnamefont {Benjamin}\ \bibnamefont {Villalonga}}, \bibinfo {author}
  {\bibfnamefont {Theodore}\ \bibnamefont {White}}, \bibinfo {author}
  {\bibfnamefont {Z.~Jamie}\ \bibnamefont {Yao}}, \bibinfo {author}
  {\bibfnamefont {Ping}\ \bibnamefont {Yeh}}, \bibinfo {author} {\bibfnamefont
  {Adam}\ \bibnamefont {Zalcman}}, \bibinfo {author} {\bibfnamefont {Hartmut}\
  \bibnamefont {Neven}}, \ and\ \bibinfo {author} {\bibfnamefont {John~M.}\
  \bibnamefont {Martinis}},\ }\bibfield  {title} {\enquote {\bibinfo {title}
  {Quantum supremacy using a programmable superconducting processor},}\ }\href
  {\doibase doi:10.1038/s41586-019-1666-5} {\bibfield  {journal} {\bibinfo
  {journal} {Nature}\ }\textbf {\bibinfo {volume} {574}},\ \bibinfo {pages}
  {505--510} (\bibinfo {year} {2019})}\BibitemShut {NoStop}%
\bibitem [{\citenamefont {Rudinger}\ \emph {et~al.}(2019)\citenamefont
  {Rudinger}, \citenamefont {Proctor}, \citenamefont {Langharst}, \citenamefont
  {Sarovar}, \citenamefont {Young},\ and\ \citenamefont
  {Blume-Kohout}}]{Rudinger2019}%
  \BibitemOpen
  \bibfield  {author} {\bibinfo {author} {\bibfnamefont {Kenneth}\ \bibnamefont
  {Rudinger}}, \bibinfo {author} {\bibfnamefont {Timothy}\ \bibnamefont
  {Proctor}}, \bibinfo {author} {\bibfnamefont {Dylan}\ \bibnamefont
  {Langharst}}, \bibinfo {author} {\bibfnamefont {Mohan}\ \bibnamefont
  {Sarovar}}, \bibinfo {author} {\bibfnamefont {Kevin}\ \bibnamefont {Young}},
  \ and\ \bibinfo {author} {\bibfnamefont {Robin}\ \bibnamefont
  {Blume-Kohout}},\ }\bibfield  {title} {\enquote {\bibinfo {title} {Probing
  context-dependent errors in quantum processors},}\ }\href {\doibase
  10.1103/PhysRevX.9.021045} {\bibfield  {journal} {\bibinfo  {journal} {Phys.
  Rev. X}\ }\textbf {\bibinfo {volume} {9}},\ \bibinfo {pages} {021045}
  (\bibinfo {year} {2019})}\BibitemShut {NoStop}%
\bibitem [{\citenamefont {Sarovar}\ \emph {et~al.}(2019)\citenamefont
  {Sarovar}, \citenamefont {Proctor}, \citenamefont {Rudinger}, \citenamefont
  {Young}, \citenamefont {Nielsen},\ and\ \citenamefont
  {Blume-Kohout}}]{Sarovar2019}%
  \BibitemOpen
  \bibfield  {author} {\bibinfo {author} {\bibfnamefont {M.}~\bibnamefont
  {Sarovar}}, \bibinfo {author} {\bibfnamefont {T.}~\bibnamefont {Proctor}},
  \bibinfo {author} {\bibfnamefont {K.}~\bibnamefont {Rudinger}}, \bibinfo
  {author} {\bibfnamefont {K.}~\bibnamefont {Young}}, \bibinfo {author}
  {\bibfnamefont {E.}~\bibnamefont {Nielsen}}, \ and\ \bibinfo {author}
  {\bibfnamefont {R.}~\bibnamefont {Blume-Kohout}},\ }\bibfield  {title}
  {\enquote {\bibinfo {title} {Detecting crosstalk errors in quantum
  information processors},}\ }\href {https://arxiv.org/abs/1908.09855}
  {\bibfield  {journal} {\bibinfo  {journal} {arXiv:1908.09855}\ } (\bibinfo
  {year} {2019})}\BibitemShut {NoStop}%
\bibitem [{\citenamefont {McKay}\ \emph {et~al.}(2020)\citenamefont {McKay},
  \citenamefont {Cross}, \citenamefont {Wood},\ and\ \citenamefont
  {Gambetta}}]{McKay2020}%
  \BibitemOpen
  \bibfield  {author} {\bibinfo {author} {\bibfnamefont {D.~C.}\ \bibnamefont
  {McKay}}, \bibinfo {author} {\bibfnamefont {A.~W.}\ \bibnamefont {Cross}},
  \bibinfo {author} {\bibfnamefont {C.~J.}\ \bibnamefont {Wood}}, \ and\
  \bibinfo {author} {\bibfnamefont {J.~M.}\ \bibnamefont {Gambetta}},\
  }\bibfield  {title} {\enquote {\bibinfo {title} {Correlated randomized
  benchmarking},}\ }\href {https://arxiv.org/abs/2003.02354} {\bibfield
  {journal} {\bibinfo  {journal} {arXiv:2003.02354}\ } (\bibinfo {year}
  {2020})}\BibitemShut {NoStop}%
\bibitem [{\citenamefont {Wright}\ \emph {et~al.}(2019)\citenamefont {Wright},
  \citenamefont {Beck}, \citenamefont {Debnath}, \citenamefont {Amini},
  \citenamefont {Nam}, \citenamefont {Grzesiak}, \citenamefont {Chen},
  \citenamefont {Pisenti}, \citenamefont {Chmielewski}, \citenamefont
  {Collins}, \citenamefont {Hudek}, \citenamefont {Mizrahi}, \citenamefont
  {Wong-Campos}, \citenamefont {Allen}, \citenamefont {Apisdorf}, \citenamefont
  {Solomon}, \citenamefont {Williams}, \citenamefont {Ducore}, \citenamefont
  {Blinov}, \citenamefont {Kreikemeier}, \citenamefont {Chaplin}, \citenamefont
  {Keesan}, \citenamefont {Monroe},\ and\ \citenamefont {Kim}}]{Wright2019}%
  \BibitemOpen
  \bibfield  {author} {\bibinfo {author} {\bibfnamefont {K.}~\bibnamefont
  {Wright}}, \bibinfo {author} {\bibfnamefont {K.~M.}\ \bibnamefont {Beck}},
  \bibinfo {author} {\bibfnamefont {S.}~\bibnamefont {Debnath}}, \bibinfo
  {author} {\bibfnamefont {J.~M.}\ \bibnamefont {Amini}}, \bibinfo {author}
  {\bibfnamefont {Y.}~\bibnamefont {Nam}}, \bibinfo {author} {\bibfnamefont
  {N.}~\bibnamefont {Grzesiak}}, \bibinfo {author} {\bibfnamefont {J.~.}\
  \bibnamefont {Chen}}, \bibinfo {author} {\bibfnamefont {N.~C.}\ \bibnamefont
  {Pisenti}}, \bibinfo {author} {\bibfnamefont {M.}~\bibnamefont
  {Chmielewski}}, \bibinfo {author} {\bibfnamefont {C.}~\bibnamefont
  {Collins}}, \bibinfo {author} {\bibfnamefont {K.~M.}\ \bibnamefont {Hudek}},
  \bibinfo {author} {\bibfnamefont {J.}~\bibnamefont {Mizrahi}}, \bibinfo
  {author} {\bibfnamefont {J.~D.}\ \bibnamefont {Wong-Campos}}, \bibinfo
  {author} {\bibfnamefont {S.}~\bibnamefont {Allen}}, \bibinfo {author}
  {\bibfnamefont {J.}~\bibnamefont {Apisdorf}}, \bibinfo {author}
  {\bibfnamefont {P.}~\bibnamefont {Solomon}}, \bibinfo {author} {\bibfnamefont
  {M.}~\bibnamefont {Williams}}, \bibinfo {author} {\bibfnamefont {A.~M.}\
  \bibnamefont {Ducore}}, \bibinfo {author} {\bibfnamefont {A.}~\bibnamefont
  {Blinov}}, \bibinfo {author} {\bibfnamefont {S.~M.}\ \bibnamefont
  {Kreikemeier}}, \bibinfo {author} {\bibfnamefont {V.}~\bibnamefont
  {Chaplin}}, \bibinfo {author} {\bibfnamefont {M.}~\bibnamefont {Keesan}},
  \bibinfo {author} {\bibfnamefont {C.}~\bibnamefont {Monroe}}, \ and\ \bibinfo
  {author} {\bibfnamefont {J.}~\bibnamefont {Kim}},\ }\bibfield  {title}
  {\enquote {\bibinfo {title} {Benchmarking an 11-qubit quantum computer},}\
  }\href {https://doi.org/10.1038/s41467-019-13534-2} {\bibfield  {journal}
  {\bibinfo  {journal} {Nature Communications}\ }\textbf {\bibinfo {volume}
  {10}},\ \bibinfo {pages} {5464} (\bibinfo {year} {2019})}\BibitemShut
  {NoStop}%
\bibitem [{\citenamefont {Erhard}\ \emph {et~al.}(2019)\citenamefont {Erhard},
  \citenamefont {Wallman}, \citenamefont {Postler}, \citenamefont {Meth},
  \citenamefont {Stricker}, \citenamefont {Martinez}, \citenamefont
  {Schindler}, \citenamefont {Monz}, \citenamefont {Emerson},\ and\
  \citenamefont {Blatt}}]{Erhard2019}%
  \BibitemOpen
  \bibfield  {author} {\bibinfo {author} {\bibfnamefont {Alexander}\
  \bibnamefont {Erhard}}, \bibinfo {author} {\bibfnamefont {Joel~J.}\
  \bibnamefont {Wallman}}, \bibinfo {author} {\bibfnamefont {Lukas}\
  \bibnamefont {Postler}}, \bibinfo {author} {\bibfnamefont {Michael}\
  \bibnamefont {Meth}}, \bibinfo {author} {\bibfnamefont {Roman}\ \bibnamefont
  {Stricker}}, \bibinfo {author} {\bibfnamefont {Esteban~A.}\ \bibnamefont
  {Martinez}}, \bibinfo {author} {\bibfnamefont {Philipp}\ \bibnamefont
  {Schindler}}, \bibinfo {author} {\bibfnamefont {Thomas}\ \bibnamefont
  {Monz}}, \bibinfo {author} {\bibfnamefont {Joseph}\ \bibnamefont {Emerson}},
  \ and\ \bibinfo {author} {\bibfnamefont {Rainer}\ \bibnamefont {Blatt}},\
  }\bibfield  {title} {\enquote {\bibinfo {title} {Characterizing large-scale
  quantum computers via cycle benchmarking},}\ }\href
  {https://doi.org/10.1038/s41467-019-13068-7} {\bibfield  {journal} {\bibinfo
  {journal} {Nature Communications}\ }\textbf {\bibinfo {volume} {10}},\
  \bibinfo {pages} {5347} (\bibinfo {year} {2019})}\BibitemShut {NoStop}%
\bibitem [{\citenamefont {Gaebler}\ \emph {et~al.}(2016)\citenamefont
  {Gaebler}, \citenamefont {Tan}, \citenamefont {Lin}, \citenamefont {Wan},
  \citenamefont {Bowler}, \citenamefont {Keith}, \citenamefont {Glancy},
  \citenamefont {Coakley}, \citenamefont {Knill}, \citenamefont {Leibfried},\
  and\ \citenamefont {et~al.}}]{Gaebler2016}%
  \BibitemOpen
  \bibfield  {author} {\bibinfo {author} {\bibfnamefont {J.P.}\ \bibnamefont
  {Gaebler}}, \bibinfo {author} {\bibfnamefont {T.R.}\ \bibnamefont {Tan}},
  \bibinfo {author} {\bibfnamefont {Y.}~\bibnamefont {Lin}}, \bibinfo {author}
  {\bibfnamefont {Y.}~\bibnamefont {Wan}}, \bibinfo {author} {\bibfnamefont
  {R.}~\bibnamefont {Bowler}}, \bibinfo {author} {\bibfnamefont {A.C.}\
  \bibnamefont {Keith}}, \bibinfo {author} {\bibfnamefont {S.}~\bibnamefont
  {Glancy}}, \bibinfo {author} {\bibfnamefont {K.}~\bibnamefont {Coakley}},
  \bibinfo {author} {\bibfnamefont {E.}~\bibnamefont {Knill}}, \bibinfo
  {author} {\bibfnamefont {D.}~\bibnamefont {Leibfried}}, \ and\ \bibinfo
  {author} {\bibnamefont {et~al.}},\ }\bibfield  {title} {\enquote {\bibinfo
  {title} {High-fidelity universal gate set for {Be9+} ion qubits},}\ }\href
  {\doibase 10.1103/physrevlett.117.060505} {\bibfield  {journal} {\bibinfo
  {journal} {Physical Review Letters}\ }\textbf {\bibinfo {volume} {117}},\
  \bibinfo {pages} {060505} (\bibinfo {year} {2016})}\BibitemShut {NoStop}%
\bibitem [{\citenamefont {Ballance}\ \emph {et~al.}(2016)\citenamefont
  {Ballance}, \citenamefont {Harty}, \citenamefont {Linke}, \citenamefont
  {Sepiol},\ and\ \citenamefont {Lucas}}]{Ballance2016}%
  \BibitemOpen
  \bibfield  {author} {\bibinfo {author} {\bibfnamefont {C.~J.}\ \bibnamefont
  {Ballance}}, \bibinfo {author} {\bibfnamefont {T.~P.}\ \bibnamefont {Harty}},
  \bibinfo {author} {\bibfnamefont {N.~M.}\ \bibnamefont {Linke}}, \bibinfo
  {author} {\bibfnamefont {M.~A.}\ \bibnamefont {Sepiol}}, \ and\ \bibinfo
  {author} {\bibfnamefont {D.~M.}\ \bibnamefont {Lucas}},\ }\bibfield  {title}
  {\enquote {\bibinfo {title} {High-fidelity quantum logic gates using
  trapped-ion hyperfine qubits},}\ }\href {\doibase
  10.1103/PhysRevLett.117.060504} {\bibfield  {journal} {\bibinfo  {journal}
  {Phys. Rev. Lett.}\ }\textbf {\bibinfo {volume} {117}},\ \bibinfo {pages}
  {060504} (\bibinfo {year} {2016})}\BibitemShut {NoStop}%
\bibitem [{\citenamefont {Blais}\ \emph {et~al.}(2004)\citenamefont {Blais},
  \citenamefont {Huang}, \citenamefont {Wallraff}, \citenamefont {Girvin},\
  and\ \citenamefont {Schoelkopf}}]{Blais2004}%
  \BibitemOpen
  \bibfield  {author} {\bibinfo {author} {\bibfnamefont {Alexandre}\
  \bibnamefont {Blais}}, \bibinfo {author} {\bibfnamefont {Ren-Shou}\
  \bibnamefont {Huang}}, \bibinfo {author} {\bibfnamefont {Andreas}\
  \bibnamefont {Wallraff}}, \bibinfo {author} {\bibfnamefont {S.~M.}\
  \bibnamefont {Girvin}}, \ and\ \bibinfo {author} {\bibfnamefont {R.~J.}\
  \bibnamefont {Schoelkopf}},\ }\bibfield  {title} {\enquote {\bibinfo {title}
  {Cavity quantum electrodynamics for superconducting electrical circuits: An
  architecture for quantum computation},}\ }\href {\doibase
  10.1103/PhysRevA.69.062320} {\bibfield  {journal} {\bibinfo  {journal} {Phys.
  Rev. A}\ }\textbf {\bibinfo {volume} {69}},\ \bibinfo {pages} {062320}
  (\bibinfo {year} {2004})}\BibitemShut {NoStop}%
\bibitem [{\citenamefont {Wallraff}\ \emph {et~al.}(2004)\citenamefont
  {Wallraff}, \citenamefont {Schuster}, \citenamefont {Blais}, \citenamefont
  {Frunzio}, \citenamefont {Huang}, \citenamefont {Majer}, \citenamefont
  {Kumar}, \citenamefont {Girvin},\ and\ \citenamefont
  {Schoelkopf}}]{Wallraff2004}%
  \BibitemOpen
  \bibfield  {author} {\bibinfo {author} {\bibfnamefont {A.}~\bibnamefont
  {Wallraff}}, \bibinfo {author} {\bibfnamefont {D.~I.}\ \bibnamefont
  {Schuster}}, \bibinfo {author} {\bibfnamefont {A.}~\bibnamefont {Blais}},
  \bibinfo {author} {\bibfnamefont {L.}~\bibnamefont {Frunzio}}, \bibinfo
  {author} {\bibfnamefont {R.-S.}\ \bibnamefont {Huang}}, \bibinfo {author}
  {\bibfnamefont {J.}~\bibnamefont {Majer}}, \bibinfo {author} {\bibfnamefont
  {S.}~\bibnamefont {Kumar}}, \bibinfo {author} {\bibfnamefont {S.~M.}\
  \bibnamefont {Girvin}}, \ and\ \bibinfo {author} {\bibfnamefont {R.~J.}\
  \bibnamefont {Schoelkopf}},\ }\bibfield  {title} {\enquote {\bibinfo {title}
  {Strong coupling of a single photon to a superconducting qubit using circuit
  quantum electrodynamics},}\ }\href {\doibase 10.1038/nature02851} {\bibfield
  {journal} {\bibinfo  {journal} {Nature}\ }\textbf {\bibinfo {volume} {431}},\
  \bibinfo {pages} {162--167} (\bibinfo {year} {2004})}\BibitemShut {NoStop}%
\bibitem [{\citenamefont {Schuster}\ \emph {et~al.}(2007)\citenamefont
  {Schuster}, \citenamefont {Houck}, \citenamefont {Schreier}, \citenamefont
  {Wallraff}, \citenamefont {Gambetta}, \citenamefont {Blais}, \citenamefont
  {Frunzio}, \citenamefont {Majer}, \citenamefont {Johnson}, \citenamefont
  {Devoret}, \citenamefont {Girvin},\ and\ \citenamefont
  {Schoelkopf}}]{Schuster2007a}%
  \BibitemOpen
  \bibfield  {author} {\bibinfo {author} {\bibfnamefont {D.~I.}\ \bibnamefont
  {Schuster}}, \bibinfo {author} {\bibfnamefont {A.~A.}\ \bibnamefont {Houck}},
  \bibinfo {author} {\bibfnamefont {J.~A.}\ \bibnamefont {Schreier}}, \bibinfo
  {author} {\bibfnamefont {A.}~\bibnamefont {Wallraff}}, \bibinfo {author}
  {\bibfnamefont {J.~M.}\ \bibnamefont {Gambetta}}, \bibinfo {author}
  {\bibfnamefont {A.}~\bibnamefont {Blais}}, \bibinfo {author} {\bibfnamefont
  {L.}~\bibnamefont {Frunzio}}, \bibinfo {author} {\bibfnamefont
  {J.}~\bibnamefont {Majer}}, \bibinfo {author} {\bibfnamefont
  {B.}~\bibnamefont {Johnson}}, \bibinfo {author} {\bibfnamefont {M.~H.}\
  \bibnamefont {Devoret}}, \bibinfo {author} {\bibfnamefont {S.~M.}\
  \bibnamefont {Girvin}}, \ and\ \bibinfo {author} {\bibfnamefont {R.~J.}\
  \bibnamefont {Schoelkopf}},\ }\bibfield  {title} {\enquote {\bibinfo {title}
  {Resolving photon number states in a superconducting circuit},}\ }\href
  {\doibase 10.1038/nature05461} {\bibfield  {journal} {\bibinfo  {journal}
  {Nature}\ }\textbf {\bibinfo {volume} {445}},\ \bibinfo {pages} {515--518}
  (\bibinfo {year} {2007})}\BibitemShut {NoStop}%
\bibitem [{\citenamefont {DiCarlo}\ \emph {et~al.}(2009)\citenamefont
  {DiCarlo}, \citenamefont {Chow}, \citenamefont {Gambetta}, \citenamefont
  {Bishop}, \citenamefont {Johnson}, \citenamefont {Schuster}, \citenamefont
  {Majer}, \citenamefont {Blais}, \citenamefont {Frunzio}, \citenamefont
  {Girvin},\ and\ \citenamefont {Schoelkopf}}]{DiCarlo2009}%
  \BibitemOpen
  \bibfield  {author} {\bibinfo {author} {\bibfnamefont {L.}~\bibnamefont
  {DiCarlo}}, \bibinfo {author} {\bibfnamefont {J.~M.}\ \bibnamefont {Chow}},
  \bibinfo {author} {\bibfnamefont {J.~M.}\ \bibnamefont {Gambetta}}, \bibinfo
  {author} {\bibfnamefont {Lev~S.}\ \bibnamefont {Bishop}}, \bibinfo {author}
  {\bibfnamefont {B.~R.}\ \bibnamefont {Johnson}}, \bibinfo {author}
  {\bibfnamefont {D.~I.}\ \bibnamefont {Schuster}}, \bibinfo {author}
  {\bibfnamefont {J.}~\bibnamefont {Majer}}, \bibinfo {author} {\bibfnamefont
  {A.}~\bibnamefont {Blais}}, \bibinfo {author} {\bibfnamefont
  {L.}~\bibnamefont {Frunzio}}, \bibinfo {author} {\bibfnamefont {S.~M.}\
  \bibnamefont {Girvin}}, \ and\ \bibinfo {author} {\bibfnamefont {R.~J.}\
  \bibnamefont {Schoelkopf}},\ }\bibfield  {title} {\enquote {\bibinfo {title}
  {Demonstration of two-qubit algorithms with a superconducting quantum
  processor},}\ }\href {\doibase 10.1038/nature08121} {\bibfield  {journal}
  {\bibinfo  {journal} {Nature}\ }\textbf {\bibinfo {volume} {460}},\ \bibinfo
  {pages} {240--244} (\bibinfo {year} {2009})}\BibitemShut {NoStop}%
\bibitem [{\citenamefont {McKay}\ \emph {et~al.}(2016)\citenamefont {McKay},
  \citenamefont {Filipp}, \citenamefont {Mezzacapo}, \citenamefont {Magesan},
  \citenamefont {Chow},\ and\ \citenamefont {Gambetta}}]{McKay2016}%
  \BibitemOpen
  \bibfield  {author} {\bibinfo {author} {\bibfnamefont {David~C.}\
  \bibnamefont {McKay}}, \bibinfo {author} {\bibfnamefont {Stefan}\
  \bibnamefont {Filipp}}, \bibinfo {author} {\bibfnamefont {Antonio}\
  \bibnamefont {Mezzacapo}}, \bibinfo {author} {\bibfnamefont {Easwar}\
  \bibnamefont {Magesan}}, \bibinfo {author} {\bibfnamefont {Jerry~M.}\
  \bibnamefont {Chow}}, \ and\ \bibinfo {author} {\bibfnamefont {Jay~M.}\
  \bibnamefont {Gambetta}},\ }\bibfield  {title} {\enquote {\bibinfo {title}
  {Universal gate for fixed-frequency qubits via a tunable bus},}\ }\href
  {\doibase 10.1103/PhysRevApplied.6.064007} {\bibfield  {journal} {\bibinfo
  {journal} {Phys. Rev. Applied}\ }\textbf {\bibinfo {volume} {6}},\ \bibinfo
  {pages} {064007} (\bibinfo {year} {2016})}\BibitemShut {NoStop}%
\bibitem [{\citenamefont {Caldwell}\ \emph {et~al.}(2018)\citenamefont
  {Caldwell}, \citenamefont {Didier}, \citenamefont {Ryan}, \citenamefont
  {Sete}, \citenamefont {Hudson}, \citenamefont {Karalekas}, \citenamefont
  {Manenti}, \citenamefont {da~Silva}, \citenamefont {Sinclair}, \citenamefont
  {Acala}, \citenamefont {Alidoust}, \citenamefont {Angeles}, \citenamefont
  {Bestwick}, \citenamefont {Block}, \citenamefont {Bloom}, \citenamefont
  {Bradley}, \citenamefont {Bui}, \citenamefont {Capelluto}, \citenamefont
  {Chilcott}, \citenamefont {Cordova}, \citenamefont {Crossman}, \citenamefont
  {Curtis}, \citenamefont {Deshpande}, \citenamefont {Bouayadi}, \citenamefont
  {Girshovich}, \citenamefont {Hong}, \citenamefont {Kuang}, \citenamefont
  {Lenihan}, \citenamefont {Manning}, \citenamefont {Marchenkov}, \citenamefont
  {Marshall}, \citenamefont {Maydra}, \citenamefont {Mohan}, \citenamefont
  {O'Brien}, \citenamefont {Osborn}, \citenamefont {Otterbach}, \citenamefont
  {Papageorge}, \citenamefont {Paquette}, \citenamefont {Pelstring},
  \citenamefont {Polloreno}, \citenamefont {Prawiroatmodjo}, \citenamefont
  {Rawat}, \citenamefont {Reagor}, \citenamefont {Renzas}, \citenamefont
  {Rubin}, \citenamefont {Russell}, \citenamefont {Rust}, \citenamefont
  {Scarabelli}, \citenamefont {Scheer}, \citenamefont {Selvanayagam},
  \citenamefont {Smith}, \citenamefont {Staley}, \citenamefont {Suska},
  \citenamefont {Tezak}, \citenamefont {Thompson}, \citenamefont {To},
  \citenamefont {Vahidpour}, \citenamefont {Vodrahalli}, \citenamefont
  {Whyland}, \citenamefont {Yadav}, \citenamefont {Zeng},\ and\ \citenamefont
  {Rigetti}}]{Caldwell2018}%
  \BibitemOpen
  \bibfield  {author} {\bibinfo {author} {\bibfnamefont {S.~A.}\ \bibnamefont
  {Caldwell}}, \bibinfo {author} {\bibfnamefont {N.}~\bibnamefont {Didier}},
  \bibinfo {author} {\bibfnamefont {C.~A.}\ \bibnamefont {Ryan}}, \bibinfo
  {author} {\bibfnamefont {E.~A.}\ \bibnamefont {Sete}}, \bibinfo {author}
  {\bibfnamefont {A.}~\bibnamefont {Hudson}}, \bibinfo {author} {\bibfnamefont
  {P.}~\bibnamefont {Karalekas}}, \bibinfo {author} {\bibfnamefont
  {R.}~\bibnamefont {Manenti}}, \bibinfo {author} {\bibfnamefont {M.~P.}\
  \bibnamefont {da~Silva}}, \bibinfo {author} {\bibfnamefont {R.}~\bibnamefont
  {Sinclair}}, \bibinfo {author} {\bibfnamefont {E.}~\bibnamefont {Acala}},
  \bibinfo {author} {\bibfnamefont {N.}~\bibnamefont {Alidoust}}, \bibinfo
  {author} {\bibfnamefont {J.}~\bibnamefont {Angeles}}, \bibinfo {author}
  {\bibfnamefont {A.}~\bibnamefont {Bestwick}}, \bibinfo {author}
  {\bibfnamefont {M.}~\bibnamefont {Block}}, \bibinfo {author} {\bibfnamefont
  {B.}~\bibnamefont {Bloom}}, \bibinfo {author} {\bibfnamefont
  {A.}~\bibnamefont {Bradley}}, \bibinfo {author} {\bibfnamefont
  {C.}~\bibnamefont {Bui}}, \bibinfo {author} {\bibfnamefont {L.}~\bibnamefont
  {Capelluto}}, \bibinfo {author} {\bibfnamefont {R.}~\bibnamefont {Chilcott}},
  \bibinfo {author} {\bibfnamefont {J.}~\bibnamefont {Cordova}}, \bibinfo
  {author} {\bibfnamefont {G.}~\bibnamefont {Crossman}}, \bibinfo {author}
  {\bibfnamefont {M.}~\bibnamefont {Curtis}}, \bibinfo {author} {\bibfnamefont
  {S.}~\bibnamefont {Deshpande}}, \bibinfo {author} {\bibfnamefont {T.~El}\
  \bibnamefont {Bouayadi}}, \bibinfo {author} {\bibfnamefont {D.}~\bibnamefont
  {Girshovich}}, \bibinfo {author} {\bibfnamefont {S.}~\bibnamefont {Hong}},
  \bibinfo {author} {\bibfnamefont {K.}~\bibnamefont {Kuang}}, \bibinfo
  {author} {\bibfnamefont {M.}~\bibnamefont {Lenihan}}, \bibinfo {author}
  {\bibfnamefont {T.}~\bibnamefont {Manning}}, \bibinfo {author} {\bibfnamefont
  {A.}~\bibnamefont {Marchenkov}}, \bibinfo {author} {\bibfnamefont
  {J.}~\bibnamefont {Marshall}}, \bibinfo {author} {\bibfnamefont
  {R.}~\bibnamefont {Maydra}}, \bibinfo {author} {\bibfnamefont
  {Y.}~\bibnamefont {Mohan}}, \bibinfo {author} {\bibfnamefont
  {W.}~\bibnamefont {O'Brien}}, \bibinfo {author} {\bibfnamefont
  {C.}~\bibnamefont {Osborn}}, \bibinfo {author} {\bibfnamefont
  {J.}~\bibnamefont {Otterbach}}, \bibinfo {author} {\bibfnamefont
  {A.}~\bibnamefont {Papageorge}}, \bibinfo {author} {\bibfnamefont {J.-P.}\
  \bibnamefont {Paquette}}, \bibinfo {author} {\bibfnamefont {M.}~\bibnamefont
  {Pelstring}}, \bibinfo {author} {\bibfnamefont {A.}~\bibnamefont
  {Polloreno}}, \bibinfo {author} {\bibfnamefont {G.}~\bibnamefont
  {Prawiroatmodjo}}, \bibinfo {author} {\bibfnamefont {V.}~\bibnamefont
  {Rawat}}, \bibinfo {author} {\bibfnamefont {M.}~\bibnamefont {Reagor}},
  \bibinfo {author} {\bibfnamefont {R.}~\bibnamefont {Renzas}}, \bibinfo
  {author} {\bibfnamefont {N.}~\bibnamefont {Rubin}}, \bibinfo {author}
  {\bibfnamefont {D.}~\bibnamefont {Russell}}, \bibinfo {author} {\bibfnamefont
  {M.}~\bibnamefont {Rust}}, \bibinfo {author} {\bibfnamefont {D.}~\bibnamefont
  {Scarabelli}}, \bibinfo {author} {\bibfnamefont {M.}~\bibnamefont {Scheer}},
  \bibinfo {author} {\bibfnamefont {M.}~\bibnamefont {Selvanayagam}}, \bibinfo
  {author} {\bibfnamefont {R.}~\bibnamefont {Smith}}, \bibinfo {author}
  {\bibfnamefont {A.}~\bibnamefont {Staley}}, \bibinfo {author} {\bibfnamefont
  {M.}~\bibnamefont {Suska}}, \bibinfo {author} {\bibfnamefont
  {N.}~\bibnamefont {Tezak}}, \bibinfo {author} {\bibfnamefont {D.~C.}\
  \bibnamefont {Thompson}}, \bibinfo {author} {\bibfnamefont {T.-W.}\
  \bibnamefont {To}}, \bibinfo {author} {\bibfnamefont {M.}~\bibnamefont
  {Vahidpour}}, \bibinfo {author} {\bibfnamefont {N.}~\bibnamefont
  {Vodrahalli}}, \bibinfo {author} {\bibfnamefont {T.}~\bibnamefont {Whyland}},
  \bibinfo {author} {\bibfnamefont {K.}~\bibnamefont {Yadav}}, \bibinfo
  {author} {\bibfnamefont {W.}~\bibnamefont {Zeng}}, \ and\ \bibinfo {author}
  {\bibfnamefont {C.}~\bibnamefont {Rigetti}},\ }\bibfield  {title} {\enquote
  {\bibinfo {title} {Parametrically activated entangling gates using transmon
  qubits},}\ }\href {\doibase 10.1103/PhysRevApplied.10.034050} {\bibfield
  {journal} {\bibinfo  {journal} {Phys. Rev. Applied}\ }\textbf {\bibinfo
  {volume} {10}},\ \bibinfo {pages} {034050} (\bibinfo {year}
  {2018})}\BibitemShut {NoStop}%
\bibitem [{\citenamefont {Wallraff}\ \emph {et~al.}(2005)\citenamefont
  {Wallraff}, \citenamefont {Schuster}, \citenamefont {Blais}, \citenamefont
  {Frunzio}, \citenamefont {Majer}, \citenamefont {Devoret}, \citenamefont
  {Girvin},\ and\ \citenamefont {Schoelkopf}}]{Wallraff2005}%
  \BibitemOpen
  \bibfield  {author} {\bibinfo {author} {\bibfnamefont {A.}~\bibnamefont
  {Wallraff}}, \bibinfo {author} {\bibfnamefont {D.~I.}\ \bibnamefont
  {Schuster}}, \bibinfo {author} {\bibfnamefont {A.}~\bibnamefont {Blais}},
  \bibinfo {author} {\bibfnamefont {L.}~\bibnamefont {Frunzio}}, \bibinfo
  {author} {\bibfnamefont {J.}~\bibnamefont {Majer}}, \bibinfo {author}
  {\bibfnamefont {M.~H.}\ \bibnamefont {Devoret}}, \bibinfo {author}
  {\bibfnamefont {S.~M.}\ \bibnamefont {Girvin}}, \ and\ \bibinfo {author}
  {\bibfnamefont {R.~J.}\ \bibnamefont {Schoelkopf}},\ }\bibfield  {title}
  {\enquote {\bibinfo {title} {Approaching unit visibility for control of a
  superconducting qubit with dispersive readout},}\ }\href {\doibase
  10.1103/PhysRevLett.95.060501} {\bibfield  {journal} {\bibinfo  {journal}
  {Phys. Rev. Lett.}\ }\textbf {\bibinfo {volume} {95}},\ \bibinfo {pages}
  {060501--4} (\bibinfo {year} {2005})}\BibitemShut {NoStop}%
\bibitem [{\citenamefont {Thompson}\ \emph {et~al.}(2008)\citenamefont
  {Thompson}, \citenamefont {Zwickl}, \citenamefont {Jayich}, \citenamefont
  {Marquardt}, \citenamefont {Girvin},\ and\ \citenamefont
  {Harris}}]{Thompson2008}%
  \BibitemOpen
  \bibfield  {author} {\bibinfo {author} {\bibfnamefont {J.~D.}\ \bibnamefont
  {Thompson}}, \bibinfo {author} {\bibfnamefont {B.~M.}\ \bibnamefont
  {Zwickl}}, \bibinfo {author} {\bibfnamefont {A.~M.}\ \bibnamefont {Jayich}},
  \bibinfo {author} {\bibfnamefont {Florian}\ \bibnamefont {Marquardt}},
  \bibinfo {author} {\bibfnamefont {S.~M.}\ \bibnamefont {Girvin}}, \ and\
  \bibinfo {author} {\bibfnamefont {J.~G.~E.}\ \bibnamefont {Harris}},\
  }\bibfield  {title} {\enquote {\bibinfo {title} {Strong dispersive coupling
  of a high-finesse cavity to a micromechanical membrane},}\ }\href
  {https://doi.org/10.1038/nature06715} {\bibfield  {journal} {\bibinfo
  {journal} {Nature}\ }\textbf {\bibinfo {volume} {452}},\ \bibinfo {pages}
  {72--75} (\bibinfo {year} {2008})}\BibitemShut {NoStop}%
\bibitem [{\citenamefont {Meineke}\ \emph {et~al.}(2012)\citenamefont
  {Meineke}, \citenamefont {Brantut}, \citenamefont {Stadler}, \citenamefont
  {Müller}, \citenamefont {Moritz},\ and\ \citenamefont
  {Esslinger}}]{Meineke2012}%
  \BibitemOpen
  \bibfield  {author} {\bibinfo {author} {\bibfnamefont {Jakob}\ \bibnamefont
  {Meineke}}, \bibinfo {author} {\bibfnamefont {Jean-Philippe}\ \bibnamefont
  {Brantut}}, \bibinfo {author} {\bibfnamefont {David}\ \bibnamefont
  {Stadler}}, \bibinfo {author} {\bibfnamefont {Torben}\ \bibnamefont
  {Müller}}, \bibinfo {author} {\bibfnamefont {Henning}\ \bibnamefont
  {Moritz}}, \ and\ \bibinfo {author} {\bibfnamefont {Tilman}\ \bibnamefont
  {Esslinger}},\ }\bibfield  {title} {\enquote {\bibinfo {title}
  {Interferometric measurement of local spin fluctuations in a quantum gas},}\
  }\href {https://doi.org/10.1038/nphys2280} {\bibfield  {journal} {\bibinfo
  {journal} {Nature Physics}\ }\textbf {\bibinfo {volume} {8}},\ \bibinfo
  {pages} {454--458} (\bibinfo {year} {2012})}\BibitemShut {NoStop}%
\bibitem [{\citenamefont {Astner}\ \emph {et~al.}(2018)\citenamefont {Astner},
  \citenamefont {Gugler}, \citenamefont {Angerer}, \citenamefont {Wald},
  \citenamefont {Putz}, \citenamefont {Mauser}, \citenamefont {Trupke},
  \citenamefont {Sumiya}, \citenamefont {Onoda}, \citenamefont {Isoya},
  \citenamefont {Schmiedmayer}, \citenamefont {Mohn},\ and\ \citenamefont
  {Majer}}]{Astner2018}%
  \BibitemOpen
  \bibfield  {author} {\bibinfo {author} {\bibfnamefont {T.}~\bibnamefont
  {Astner}}, \bibinfo {author} {\bibfnamefont {J.}~\bibnamefont {Gugler}},
  \bibinfo {author} {\bibfnamefont {A.}~\bibnamefont {Angerer}}, \bibinfo
  {author} {\bibfnamefont {S.}~\bibnamefont {Wald}}, \bibinfo {author}
  {\bibfnamefont {S.}~\bibnamefont {Putz}}, \bibinfo {author} {\bibfnamefont
  {N.~J.}\ \bibnamefont {Mauser}}, \bibinfo {author} {\bibfnamefont
  {M.}~\bibnamefont {Trupke}}, \bibinfo {author} {\bibfnamefont
  {H.}~\bibnamefont {Sumiya}}, \bibinfo {author} {\bibfnamefont
  {S.}~\bibnamefont {Onoda}}, \bibinfo {author} {\bibfnamefont
  {J.}~\bibnamefont {Isoya}}, \bibinfo {author} {\bibfnamefont
  {J.}~\bibnamefont {Schmiedmayer}}, \bibinfo {author} {\bibfnamefont
  {P.}~\bibnamefont {Mohn}}, \ and\ \bibinfo {author} {\bibfnamefont
  {J.}~\bibnamefont {Majer}},\ }\bibfield  {title} {\enquote {\bibinfo {title}
  {Solid-state electron spin lifetime limited by phononic vacuum modes},}\
  }\href {\doibase 10.1038/s41563-017-0008-y} {\bibfield  {journal} {\bibinfo
  {journal} {Nature Materials}\ } (\bibinfo {year} {2018}),\
  10.1038/s41563-017-0008-y}\BibitemShut {NoStop}%
\bibitem [{\citenamefont {Scarlino}\ \emph {et~al.}(2019)\citenamefont
  {Scarlino}, \citenamefont {van Woerkom}, \citenamefont {Stockklauser},
  \citenamefont {Koski}, \citenamefont {Collodo}, \citenamefont {Gasparinetti},
  \citenamefont {Reichl}, \citenamefont {Wegscheider}, \citenamefont {Ihn},
  \citenamefont {Ensslin},\ and\ \citenamefont {Wallraff}}]{Scarlino2019}%
  \BibitemOpen
  \bibfield  {author} {\bibinfo {author} {\bibfnamefont {P.}~\bibnamefont
  {Scarlino}}, \bibinfo {author} {\bibfnamefont {D.~J.}\ \bibnamefont {van
  Woerkom}}, \bibinfo {author} {\bibfnamefont {A.}~\bibnamefont
  {Stockklauser}}, \bibinfo {author} {\bibfnamefont {J.~V.}\ \bibnamefont
  {Koski}}, \bibinfo {author} {\bibfnamefont {M.~C.}\ \bibnamefont {Collodo}},
  \bibinfo {author} {\bibfnamefont {S.}~\bibnamefont {Gasparinetti}}, \bibinfo
  {author} {\bibfnamefont {C.}~\bibnamefont {Reichl}}, \bibinfo {author}
  {\bibfnamefont {W.}~\bibnamefont {Wegscheider}}, \bibinfo {author}
  {\bibfnamefont {T.}~\bibnamefont {Ihn}}, \bibinfo {author} {\bibfnamefont
  {K.}~\bibnamefont {Ensslin}}, \ and\ \bibinfo {author} {\bibfnamefont
  {A.}~\bibnamefont {Wallraff}},\ }\bibfield  {title} {\enquote {\bibinfo
  {title} {All-microwave control and dispersive readout of gate-defined quantum
  dot qubits in circuit quantum electrodynamics},}\ }\href {\doibase
  10.1103/PhysRevLett.122.206802} {\bibfield  {journal} {\bibinfo  {journal}
  {Phys. Rev. Lett.}\ }\textbf {\bibinfo {volume} {122}},\ \bibinfo {pages}
  {206802} (\bibinfo {year} {2019})}\BibitemShut {NoStop}%
\bibitem [{\citenamefont {Zheng}\ \emph {et~al.}(2019)\citenamefont {Zheng},
  \citenamefont {Samkharadze}, \citenamefont {Noordam}, \citenamefont {Kalhor},
  \citenamefont {Brousse}, \citenamefont {Sammak}, \citenamefont {Scappucci},\
  and\ \citenamefont {Vandersypen}}]{Zheng2019}%
  \BibitemOpen
  \bibfield  {author} {\bibinfo {author} {\bibfnamefont {Guoji}\ \bibnamefont
  {Zheng}}, \bibinfo {author} {\bibfnamefont {Nodar}\ \bibnamefont
  {Samkharadze}}, \bibinfo {author} {\bibfnamefont {Marc~L.}\ \bibnamefont
  {Noordam}}, \bibinfo {author} {\bibfnamefont {Nima}\ \bibnamefont {Kalhor}},
  \bibinfo {author} {\bibfnamefont {Delphine}\ \bibnamefont {Brousse}},
  \bibinfo {author} {\bibfnamefont {Amir}\ \bibnamefont {Sammak}}, \bibinfo
  {author} {\bibfnamefont {Giordano}\ \bibnamefont {Scappucci}}, \ and\
  \bibinfo {author} {\bibfnamefont {Lieven M.~K.}\ \bibnamefont
  {Vandersypen}},\ }\bibfield  {title} {\enquote {\bibinfo {title} {Rapid
  gate-based spin read-out in silicon using an on-chip resonator},}\ }\href
  {https://doi.org/10.1038/s41565-019-0488-9} {\bibfield  {journal} {\bibinfo
  {journal} {Nature Nanotechnology}\ }\textbf {\bibinfo {volume} {14}},\
  \bibinfo {pages} {742--746} (\bibinfo {year} {2019})}\BibitemShut {NoStop}%
\bibitem [{\citenamefont {Guti\'errez}\ \emph {et~al.}(2016)\citenamefont
  {Guti\'errez}, \citenamefont {Smith}, \citenamefont {Lulushi}, \citenamefont
  {Janardan},\ and\ \citenamefont {Brown}}]{Gutierrez2016}%
  \BibitemOpen
  \bibfield  {author} {\bibinfo {author} {\bibfnamefont {Mauricio}\
  \bibnamefont {Guti\'errez}}, \bibinfo {author} {\bibfnamefont {Conor}\
  \bibnamefont {Smith}}, \bibinfo {author} {\bibfnamefont {Livia}\ \bibnamefont
  {Lulushi}}, \bibinfo {author} {\bibfnamefont {Smitha}\ \bibnamefont
  {Janardan}}, \ and\ \bibinfo {author} {\bibfnamefont {Kenneth~R.}\
  \bibnamefont {Brown}},\ }\bibfield  {title} {\enquote {\bibinfo {title}
  {Errors and pseudothresholds for incoherent and coherent noise},}\ }\href
  {\doibase 10.1103/PhysRevA.94.042338} {\bibfield  {journal} {\bibinfo
  {journal} {Phys. Rev. A}\ }\textbf {\bibinfo {volume} {94}},\ \bibinfo
  {pages} {042338} (\bibinfo {year} {2016})}\BibitemShut {NoStop}%
\bibitem [{\citenamefont {Greenbaum}\ and\ \citenamefont
  {Dutton}(2018)}]{Greenbaum2018}%
  \BibitemOpen
  \bibfield  {author} {\bibinfo {author} {\bibfnamefont {Daniel}\ \bibnamefont
  {Greenbaum}}\ and\ \bibinfo {author} {\bibfnamefont {Zachary}\ \bibnamefont
  {Dutton}},\ }\bibfield  {title} {\enquote {\bibinfo {title} {Modeling
  coherent errors in quantum error correction},}\ }\href
  {http://stacks.iop.org/2058-9565/3/i=1/a=015007} {\bibfield  {journal}
  {\bibinfo  {journal} {Quantum Science and Technology}\ }\textbf {\bibinfo
  {volume} {3}},\ \bibinfo {pages} {015007} (\bibinfo {year}
  {2018})}\BibitemShut {NoStop}%
\bibitem [{\citenamefont {Bravyi}\ \emph {et~al.}(2018)\citenamefont {Bravyi},
  \citenamefont {Englbrecht}, \citenamefont {König},\ and\ \citenamefont
  {Peard}}]{Bravyi2018}%
  \BibitemOpen
  \bibfield  {author} {\bibinfo {author} {\bibfnamefont {Sergey}\ \bibnamefont
  {Bravyi}}, \bibinfo {author} {\bibfnamefont {Matthias}\ \bibnamefont
  {Englbrecht}}, \bibinfo {author} {\bibfnamefont {Robert}\ \bibnamefont
  {König}}, \ and\ \bibinfo {author} {\bibfnamefont {Nolan}\ \bibnamefont
  {Peard}},\ }\bibfield  {title} {\enquote {\bibinfo {title} {Correcting
  coherent errors with surface codes},}\ }\href
  {https://doi.org/10.1038/s41534-018-0106-y} {\bibfield  {journal} {\bibinfo
  {journal} {npj Quantum Information}\ }\textbf {\bibinfo {volume} {4}},\
  \bibinfo {pages} {55} (\bibinfo {year} {2018})}\BibitemShut {NoStop}%
\bibitem [{\citenamefont {Beale}\ \emph {et~al.}(2018)\citenamefont {Beale},
  \citenamefont {Wallman}, \citenamefont {Guti\'errez}, \citenamefont {Brown},\
  and\ \citenamefont {Laflamme}}]{Beale2018}%
  \BibitemOpen
  \bibfield  {author} {\bibinfo {author} {\bibfnamefont {Stefanie~J.}\
  \bibnamefont {Beale}}, \bibinfo {author} {\bibfnamefont {Joel~J.}\
  \bibnamefont {Wallman}}, \bibinfo {author} {\bibfnamefont {Mauricio}\
  \bibnamefont {Guti\'errez}}, \bibinfo {author} {\bibfnamefont {Kenneth~R.}\
  \bibnamefont {Brown}}, \ and\ \bibinfo {author} {\bibfnamefont {Raymond}\
  \bibnamefont {Laflamme}},\ }\bibfield  {title} {\enquote {\bibinfo {title}
  {Quantum error correction decoheres noise},}\ }\href {\doibase
  10.1103/PhysRevLett.121.190501} {\bibfield  {journal} {\bibinfo  {journal}
  {Phys. Rev. Lett.}\ }\textbf {\bibinfo {volume} {121}},\ \bibinfo {pages}
  {190501} (\bibinfo {year} {2018})}\BibitemShut {NoStop}%
\bibitem [{\citenamefont {Baireuther}\ \emph {et~al.}(2018)\citenamefont
  {Baireuther}, \citenamefont {O'Brien}, \citenamefont {Tarasinski},\ and\
  \citenamefont {Beenakker}}]{Baireuther2018}%
  \BibitemOpen
  \bibfield  {author} {\bibinfo {author} {\bibfnamefont {Paul}\ \bibnamefont
  {Baireuther}}, \bibinfo {author} {\bibfnamefont {Thomas~E.}\ \bibnamefont
  {O'Brien}}, \bibinfo {author} {\bibfnamefont {Brian}\ \bibnamefont
  {Tarasinski}}, \ and\ \bibinfo {author} {\bibfnamefont {Carlo W.~J.}\
  \bibnamefont {Beenakker}},\ }\bibfield  {title} {\enquote {\bibinfo {title}
  {Machine-learning-assisted correction of correlated qubit errors in a
  topological code},}\ }\href {\doibase 10.22331/q-2018-01-29-48} {\bibfield
  {journal} {\bibinfo  {journal} {{Quantum}}\ }\textbf {\bibinfo {volume}
  {2}},\ \bibinfo {pages} {48} (\bibinfo {year} {2018})}\BibitemShut {NoStop}%
\bibitem [{\citenamefont {Maskara}\ \emph {et~al.}(2019)\citenamefont
  {Maskara}, \citenamefont {Kubica},\ and\ \citenamefont
  {Jochym-O'Connor}}]{Maskara2019}%
  \BibitemOpen
  \bibfield  {author} {\bibinfo {author} {\bibfnamefont {Nishad}\ \bibnamefont
  {Maskara}}, \bibinfo {author} {\bibfnamefont {Aleksander}\ \bibnamefont
  {Kubica}}, \ and\ \bibinfo {author} {\bibfnamefont {Tomas}\ \bibnamefont
  {Jochym-O'Connor}},\ }\bibfield  {title} {\enquote {\bibinfo {title}
  {Advantages of versatile neural-network decoding for topological codes},}\
  }\href {\doibase 10.1103/PhysRevA.99.052351} {\bibfield  {journal} {\bibinfo
  {journal} {Phys. Rev. A}\ }\textbf {\bibinfo {volume} {99}},\ \bibinfo
  {pages} {052351} (\bibinfo {year} {2019})}\BibitemShut {NoStop}%
\bibitem [{\citenamefont {Viola}\ and\ \citenamefont
  {Lloyd}(1998)}]{Viola1998}%
  \BibitemOpen
  \bibfield  {author} {\bibinfo {author} {\bibfnamefont {Lorenza}\ \bibnamefont
  {Viola}}\ and\ \bibinfo {author} {\bibfnamefont {Seth}\ \bibnamefont
  {Lloyd}},\ }\bibfield  {title} {\enquote {\bibinfo {title} {Dynamical
  suppression of decoherence in two-state quantum systems},}\ }\href {\doibase
  10.1103/PhysRevA.58.2733} {\bibfield  {journal} {\bibinfo  {journal} {Phys.
  Rev. A}\ }\textbf {\bibinfo {volume} {58}},\ \bibinfo {pages} {2733--2744}
  (\bibinfo {year} {1998})}\BibitemShut {NoStop}%
\bibitem [{\citenamefont {Vandersypen}\ and\ \citenamefont
  {Chuang}(2004)}]{Vandersypen2004}%
  \BibitemOpen
  \bibfield  {author} {\bibinfo {author} {\bibfnamefont {L.~M.~K.}\
  \bibnamefont {Vandersypen}}\ and\ \bibinfo {author} {\bibfnamefont {I.~L.}\
  \bibnamefont {Chuang}},\ }\bibfield  {title} {\enquote {\bibinfo {title}
  {N{M}{R} techniques for quantum control and computation},}\ }\href {\doibase
  10.1103/RevModPhys.76.1037} {\bibfield  {journal} {\bibinfo  {journal} {Rev.
  Mod. Phys.}\ }\textbf {\bibinfo {volume} {76}},\ \bibinfo {pages} {1037}
  (\bibinfo {year} {2004})}\BibitemShut {NoStop}%
\bibitem [{\citenamefont {Bylander}\ \emph {et~al.}(2011)\citenamefont
  {Bylander}, \citenamefont {Gustavsson}, \citenamefont {Yan}, \citenamefont
  {Yoshihara}, \citenamefont {Harrabi}, \citenamefont {Fitch}, \citenamefont
  {Cory}, \citenamefont {Nakamura}, \citenamefont {Tsai},\ and\ \citenamefont
  {D.}}]{Bylander2011}%
  \BibitemOpen
  \bibfield  {author} {\bibinfo {author} {\bibfnamefont {J.}~\bibnamefont
  {Bylander}}, \bibinfo {author} {\bibfnamefont {S.}~\bibnamefont
  {Gustavsson}}, \bibinfo {author} {\bibfnamefont {F.}~\bibnamefont {Yan}},
  \bibinfo {author} {\bibfnamefont {F.}~\bibnamefont {Yoshihara}}, \bibinfo
  {author} {\bibfnamefont {K.}~\bibnamefont {Harrabi}}, \bibinfo {author}
  {\bibfnamefont {G.}~\bibnamefont {Fitch}}, \bibinfo {author} {\bibfnamefont
  {D.~G.}\ \bibnamefont {Cory}}, \bibinfo {author} {\bibfnamefont
  {Y.}~\bibnamefont {Nakamura}}, \bibinfo {author} {\bibfnamefont {J.-S.}\
  \bibnamefont {Tsai}}, \ and\ \bibinfo {author} {\bibfnamefont {Oliver~W.}\
  \bibnamefont {D.}},\ }\bibfield  {title} {\enquote {\bibinfo {title} {Noise
  spectroscopy through dynamical decoupling with a superconducting flux
  qubit},}\ }\href {\doibase doi:10.1038/nphys1994} {\bibfield  {journal}
  {\bibinfo  {journal} {Nat. Phys.}\ }\textbf {\bibinfo {volume} {7}},\
  \bibinfo {pages} {565--570} (\bibinfo {year} {2011})}\BibitemShut {NoStop}%
\bibitem [{\citenamefont {Guo}\ \emph {et~al.}(2018)\citenamefont {Guo},
  \citenamefont {Zheng}, \citenamefont {Wang}, \citenamefont {Song},
  \citenamefont {Zhang}, \citenamefont {Li}, \citenamefont {Liu}, \citenamefont
  {Deng}, \citenamefont {Huang}, \citenamefont {Zheng}, \citenamefont {Zhu},
  \citenamefont {Wang}, \citenamefont {Lu},\ and\ \citenamefont
  {Pan}}]{Guo2018a}%
  \BibitemOpen
  \bibfield  {author} {\bibinfo {author} {\bibfnamefont {Qiujiang}\
  \bibnamefont {Guo}}, \bibinfo {author} {\bibfnamefont {Shi-Biao}\
  \bibnamefont {Zheng}}, \bibinfo {author} {\bibfnamefont {Jianwen}\
  \bibnamefont {Wang}}, \bibinfo {author} {\bibfnamefont {Chao}\ \bibnamefont
  {Song}}, \bibinfo {author} {\bibfnamefont {Pengfei}\ \bibnamefont {Zhang}},
  \bibinfo {author} {\bibfnamefont {Kemin}\ \bibnamefont {Li}}, \bibinfo
  {author} {\bibfnamefont {Wuxin}\ \bibnamefont {Liu}}, \bibinfo {author}
  {\bibfnamefont {Hui}\ \bibnamefont {Deng}}, \bibinfo {author} {\bibfnamefont
  {Keqiang}\ \bibnamefont {Huang}}, \bibinfo {author} {\bibfnamefont
  {Dongning}\ \bibnamefont {Zheng}}, \bibinfo {author} {\bibfnamefont {Xiaobo}\
  \bibnamefont {Zhu}}, \bibinfo {author} {\bibfnamefont {H.}~\bibnamefont
  {Wang}}, \bibinfo {author} {\bibfnamefont {C.-Y.}\ \bibnamefont {Lu}}, \ and\
  \bibinfo {author} {\bibfnamefont {Jian-Wei}\ \bibnamefont {Pan}},\ }\bibfield
   {title} {\enquote {\bibinfo {title} {Dephasing-insensitive quantum
  information storage and processing with superconducting qubits},}\ }\href
  {\doibase 10.1103/PhysRevLett.121.130501} {\bibfield  {journal} {\bibinfo
  {journal} {Phys. Rev. Lett.}\ }\textbf {\bibinfo {volume} {121}},\ \bibinfo
  {pages} {130501} (\bibinfo {year} {2018})}\BibitemShut {NoStop}%
\bibitem [{\citenamefont {McKay}\ \emph {et~al.}(2015)\citenamefont {McKay},
  \citenamefont {Naik}, \citenamefont {Reinhold}, \citenamefont {Bishop},\ and\
  \citenamefont {Schuster}}]{McKay2015}%
  \BibitemOpen
  \bibfield  {author} {\bibinfo {author} {\bibfnamefont {David~C.}\
  \bibnamefont {McKay}}, \bibinfo {author} {\bibfnamefont {Ravi}\ \bibnamefont
  {Naik}}, \bibinfo {author} {\bibfnamefont {Philip}\ \bibnamefont {Reinhold}},
  \bibinfo {author} {\bibfnamefont {Lev~S.}\ \bibnamefont {Bishop}}, \ and\
  \bibinfo {author} {\bibfnamefont {David~I.}\ \bibnamefont {Schuster}},\
  }\bibfield  {title} {\enquote {\bibinfo {title} {High-contrast qubit
  interactions using multimode cavity qed},}\ }\href {\doibase
  10.1103/PhysRevLett.114.080501} {\bibfield  {journal} {\bibinfo  {journal}
  {Phys. Rev. Lett.}\ }\textbf {\bibinfo {volume} {114}},\ \bibinfo {pages}
  {080501} (\bibinfo {year} {2015})}\BibitemShut {NoStop}%
\bibitem [{\citenamefont {Chen}\ \emph {et~al.}(2014)\citenamefont {Chen},
  \citenamefont {Neill}, \citenamefont {Roushan}, \citenamefont {Leung},
  \citenamefont {Fang}, \citenamefont {Barends}, \citenamefont {Kelly},
  \citenamefont {Campbell}, \citenamefont {Chen}, \citenamefont {Chiaro},
  \citenamefont {Dunsworth}, \citenamefont {Jeffrey}, \citenamefont {Megrant},
  \citenamefont {Mutus}, \citenamefont {O'Malley}, \citenamefont {Quintana},
  \citenamefont {Sank}, \citenamefont {Vainsencher}, \citenamefont {Wenner},
  \citenamefont {White}, \citenamefont {Geller}, \citenamefont {Cleland},\ and\
  \citenamefont {Martinis}}]{Chen2014m}%
  \BibitemOpen
  \bibfield  {author} {\bibinfo {author} {\bibfnamefont {Yu}~\bibnamefont
  {Chen}}, \bibinfo {author} {\bibfnamefont {C.}~\bibnamefont {Neill}},
  \bibinfo {author} {\bibfnamefont {P.}~\bibnamefont {Roushan}}, \bibinfo
  {author} {\bibfnamefont {N.}~\bibnamefont {Leung}}, \bibinfo {author}
  {\bibfnamefont {M.}~\bibnamefont {Fang}}, \bibinfo {author} {\bibfnamefont
  {R.}~\bibnamefont {Barends}}, \bibinfo {author} {\bibfnamefont
  {J.}~\bibnamefont {Kelly}}, \bibinfo {author} {\bibfnamefont
  {B.}~\bibnamefont {Campbell}}, \bibinfo {author} {\bibfnamefont
  {Z.}~\bibnamefont {Chen}}, \bibinfo {author} {\bibfnamefont {B.}~\bibnamefont
  {Chiaro}}, \bibinfo {author} {\bibfnamefont {A.}~\bibnamefont {Dunsworth}},
  \bibinfo {author} {\bibfnamefont {E.}~\bibnamefont {Jeffrey}}, \bibinfo
  {author} {\bibfnamefont {A.}~\bibnamefont {Megrant}}, \bibinfo {author}
  {\bibfnamefont {J.~Y.}\ \bibnamefont {Mutus}}, \bibinfo {author}
  {\bibfnamefont {P.~J.~J.}\ \bibnamefont {O'Malley}}, \bibinfo {author}
  {\bibfnamefont {C.~M.}\ \bibnamefont {Quintana}}, \bibinfo {author}
  {\bibfnamefont {D.}~\bibnamefont {Sank}}, \bibinfo {author} {\bibfnamefont
  {A.}~\bibnamefont {Vainsencher}}, \bibinfo {author} {\bibfnamefont
  {J.}~\bibnamefont {Wenner}}, \bibinfo {author} {\bibfnamefont {T.~C.}\
  \bibnamefont {White}}, \bibinfo {author} {\bibfnamefont {Michael~R.}\
  \bibnamefont {Geller}}, \bibinfo {author} {\bibfnamefont {A.~N.}\
  \bibnamefont {Cleland}}, \ and\ \bibinfo {author} {\bibfnamefont {John~M.}\
  \bibnamefont {Martinis}},\ }\bibfield  {title} {\enquote {\bibinfo {title}
  {Qubit architecture with high coherence and fast tunable coupling},}\ }\href
  {\doibase 10.1103/PhysRevLett.113.220502} {\bibfield  {journal} {\bibinfo
  {journal} {Phys. Rev. Lett.}\ }\textbf {\bibinfo {volume} {113}},\ \bibinfo
  {pages} {220502} (\bibinfo {year} {2014})}\BibitemShut {NoStop}%
\bibitem [{\citenamefont {Yan}\ \emph {et~al.}(2018)\citenamefont {Yan},
  \citenamefont {Krantz}, \citenamefont {Sung}, \citenamefont {Kjaergaard},
  \citenamefont {Campbell}, \citenamefont {Orlando}, \citenamefont
  {Gustavsson},\ and\ \citenamefont {Oliver}}]{Yan2018b}%
  \BibitemOpen
  \bibfield  {author} {\bibinfo {author} {\bibfnamefont {Fei}\ \bibnamefont
  {Yan}}, \bibinfo {author} {\bibfnamefont {Philip}\ \bibnamefont {Krantz}},
  \bibinfo {author} {\bibfnamefont {Youngkyu}\ \bibnamefont {Sung}}, \bibinfo
  {author} {\bibfnamefont {Morten}\ \bibnamefont {Kjaergaard}}, \bibinfo
  {author} {\bibfnamefont {Daniel~L.}\ \bibnamefont {Campbell}}, \bibinfo
  {author} {\bibfnamefont {Terry~P.}\ \bibnamefont {Orlando}}, \bibinfo
  {author} {\bibfnamefont {Simon}\ \bibnamefont {Gustavsson}}, \ and\ \bibinfo
  {author} {\bibfnamefont {William~D.}\ \bibnamefont {Oliver}},\ }\bibfield
  {title} {\enquote {\bibinfo {title} {Tunable coupling scheme for implementing
  high-fidelity two-qubit gates},}\ }\href {\doibase
  10.1103/PhysRevApplied.10.054062} {\bibfield  {journal} {\bibinfo  {journal}
  {Phys. Rev. Applied}\ }\textbf {\bibinfo {volume} {10}},\ \bibinfo {pages}
  {054062} (\bibinfo {year} {2018})}\BibitemShut {NoStop}%
\bibitem [{\citenamefont {Mundada}\ \emph {et~al.}(2019)\citenamefont
  {Mundada}, \citenamefont {Zhang}, \citenamefont {Hazard},\ and\ \citenamefont
  {Houck}}]{Mundada2019}%
  \BibitemOpen
  \bibfield  {author} {\bibinfo {author} {\bibfnamefont {Pranav}\ \bibnamefont
  {Mundada}}, \bibinfo {author} {\bibfnamefont {Gengyan}\ \bibnamefont
  {Zhang}}, \bibinfo {author} {\bibfnamefont {Thomas}\ \bibnamefont {Hazard}},
  \ and\ \bibinfo {author} {\bibfnamefont {Andrew}\ \bibnamefont {Houck}},\
  }\bibfield  {title} {\enquote {\bibinfo {title} {Suppression of qubit
  crosstalk in a tunable coupling superconducting circuit},}\ }\href {\doibase
  10.1103/PhysRevApplied.12.054023} {\bibfield  {journal} {\bibinfo  {journal}
  {Phys. Rev. Applied}\ }\textbf {\bibinfo {volume} {12}},\ \bibinfo {pages}
  {054023} (\bibinfo {year} {2019})}\BibitemShut {NoStop}%
\bibitem [{\citenamefont {Li}\ \emph {et~al.}(2019)\citenamefont {Li},
  \citenamefont {Cai}, \citenamefont {Yan}, \citenamefont {Wang}, \citenamefont
  {Pan}, \citenamefont {Ma}, \citenamefont {Cai}, \citenamefont {Han},
  \citenamefont {Hua}, \citenamefont {Han}, \citenamefont {Wu}, \citenamefont
  {Zhang}, \citenamefont {Wang}, \citenamefont {Song}, \citenamefont {Duan},\
  and\ \citenamefont {Sun}}]{Li2019t}%
  \BibitemOpen
  \bibfield  {author} {\bibinfo {author} {\bibfnamefont {X.}~\bibnamefont
  {Li}}, \bibinfo {author} {\bibfnamefont {T.}~\bibnamefont {Cai}}, \bibinfo
  {author} {\bibfnamefont {H.}~\bibnamefont {Yan}}, \bibinfo {author}
  {\bibfnamefont {Z.}~\bibnamefont {Wang}}, \bibinfo {author} {\bibfnamefont
  {X.}~\bibnamefont {Pan}}, \bibinfo {author} {\bibfnamefont {Y.}~\bibnamefont
  {Ma}}, \bibinfo {author} {\bibfnamefont {W.}~\bibnamefont {Cai}}, \bibinfo
  {author} {\bibfnamefont {J.}~\bibnamefont {Han}}, \bibinfo {author}
  {\bibfnamefont {Z.}~\bibnamefont {Hua}}, \bibinfo {author} {\bibfnamefont
  {X.}~\bibnamefont {Han}}, \bibinfo {author} {\bibfnamefont {Y.}~\bibnamefont
  {Wu}}, \bibinfo {author} {\bibfnamefont {H.}~\bibnamefont {Zhang}}, \bibinfo
  {author} {\bibfnamefont {H.}~\bibnamefont {Wang}}, \bibinfo {author}
  {\bibfnamefont {Y.}~\bibnamefont {Song}}, \bibinfo {author} {\bibfnamefont
  {L.}~\bibnamefont {Duan}}, \ and\ \bibinfo {author} {\bibfnamefont
  {L.}~\bibnamefont {Sun}},\ }\bibfield  {title} {\enquote {\bibinfo {title} {A
  tunable coupler for suppressing adjacent superconducting qubit coupling},}\
  }\href {https://arxiv.org/abs/1912.10721} {\bibfield  {journal} {\bibinfo
  {journal} {arXiv:1912.10721}\ } (\bibinfo {year} {2019})}\BibitemShut
  {NoStop}%
\bibitem [{\citenamefont {Steffen}\ \emph {et~al.}(2013)\citenamefont
  {Steffen}, \citenamefont {Salathe}, \citenamefont {Oppliger}, \citenamefont
  {Kurpiers}, \citenamefont {Baur}, \citenamefont {Lang}, \citenamefont
  {Eichler}, \citenamefont {Puebla-Hellmann}, \citenamefont {Fedorov},\ and\
  \citenamefont {Wallraff}}]{Steffen2013}%
  \BibitemOpen
  \bibfield  {author} {\bibinfo {author} {\bibfnamefont {L.}~\bibnamefont
  {Steffen}}, \bibinfo {author} {\bibfnamefont {Y.}~\bibnamefont {Salathe}},
  \bibinfo {author} {\bibfnamefont {M.}~\bibnamefont {Oppliger}}, \bibinfo
  {author} {\bibfnamefont {P.}~\bibnamefont {Kurpiers}}, \bibinfo {author}
  {\bibfnamefont {M.}~\bibnamefont {Baur}}, \bibinfo {author} {\bibfnamefont
  {C.}~\bibnamefont {Lang}}, \bibinfo {author} {\bibfnamefont {C.}~\bibnamefont
  {Eichler}}, \bibinfo {author} {\bibfnamefont {G.}~\bibnamefont
  {Puebla-Hellmann}}, \bibinfo {author} {\bibfnamefont {A.}~\bibnamefont
  {Fedorov}}, \ and\ \bibinfo {author} {\bibfnamefont {A.}~\bibnamefont
  {Wallraff}},\ }\bibfield  {title} {\enquote {\bibinfo {title} {Deterministic
  quantum teleportation with feed-forward in a solid state system},}\ }\href
  {\doibase 10.1038/nature12422} {\bibfield  {journal} {\bibinfo  {journal}
  {Nature}\ }\textbf {\bibinfo {volume} {500}},\ \bibinfo {pages} {319--322}
  (\bibinfo {year} {2013})}\BibitemShut {NoStop}%
\bibitem [{\citenamefont {Takita}\ \emph {et~al.}(2016)\citenamefont {Takita},
  \citenamefont {C\'{o}rcoles}, \citenamefont {Magesan}, \citenamefont {Abdo},
  \citenamefont {Brink}, \citenamefont {Cross}, \citenamefont {Chow},\ and\
  \citenamefont {Gambetta}}]{Takita2016}%
  \BibitemOpen
  \bibfield  {author} {\bibinfo {author} {\bibfnamefont {Maika}\ \bibnamefont
  {Takita}}, \bibinfo {author} {\bibfnamefont {A.D.}\ \bibnamefont
  {C\'{o}rcoles}}, \bibinfo {author} {\bibfnamefont {Easwar}\ \bibnamefont
  {Magesan}}, \bibinfo {author} {\bibfnamefont {Baleegh}\ \bibnamefont {Abdo}},
  \bibinfo {author} {\bibfnamefont {Markus}\ \bibnamefont {Brink}}, \bibinfo
  {author} {\bibfnamefont {Andrew}\ \bibnamefont {Cross}}, \bibinfo {author}
  {\bibfnamefont {Jerry~M.}\ \bibnamefont {Chow}}, \ and\ \bibinfo {author}
  {\bibfnamefont {Jay~M.}\ \bibnamefont {Gambetta}},\ }\bibfield  {title}
  {\enquote {\bibinfo {title} {Demonstration of weight-four parity measurements
  in the surface code architecture},}\ }\href {\doibase
  10.1103/physrevlett.117.210505} {\bibfield  {journal} {\bibinfo  {journal}
  {Phys. Rev. Lett.}\ }\textbf {\bibinfo {volume} {117}},\ \bibinfo {pages}
  {210505} (\bibinfo {year} {2016})}\BibitemShut {NoStop}%
\bibitem [{\citenamefont {Andersen}\ \emph
  {et~al.}(2019{\natexlab{a}})\citenamefont {Andersen}, \citenamefont {Remm},
  \citenamefont {Lazar}, \citenamefont {Krinner}, \citenamefont {Heinsoo},
  \citenamefont {Besse}, \citenamefont {Gabureac}, \citenamefont {Wallraff},\
  and\ \citenamefont {Eichler}}]{Andersen2019}%
  \BibitemOpen
  \bibfield  {author} {\bibinfo {author} {\bibfnamefont {Christian~Kraglund}\
  \bibnamefont {Andersen}}, \bibinfo {author} {\bibfnamefont {Ants}\
  \bibnamefont {Remm}}, \bibinfo {author} {\bibfnamefont {Stefania}\
  \bibnamefont {Lazar}}, \bibinfo {author} {\bibfnamefont {Sebastian}\
  \bibnamefont {Krinner}}, \bibinfo {author} {\bibfnamefont {Johannes}\
  \bibnamefont {Heinsoo}}, \bibinfo {author} {\bibfnamefont {Jean-Claude}\
  \bibnamefont {Besse}}, \bibinfo {author} {\bibfnamefont {Mihai}\ \bibnamefont
  {Gabureac}}, \bibinfo {author} {\bibfnamefont {Andreas}\ \bibnamefont
  {Wallraff}}, \ and\ \bibinfo {author} {\bibfnamefont {Christopher}\
  \bibnamefont {Eichler}},\ }\bibfield  {title} {\enquote {\bibinfo {title}
  {Entanglement stabilization using ancilla-based parity detection and
  real-time feedback in superconducting circuits},}\ }\href
  {https://doi.org/10.1038/s41534-019-0185-4} {\bibfield  {journal} {\bibinfo
  {journal} {npj Quantum Information}\ }\textbf {\bibinfo {volume} {5}},\
  \bibinfo {pages} {69} (\bibinfo {year} {2019}{\natexlab{a}})},\ \Eprint
  {http://arxiv.org/abs/1902.06946} {arXiv:1902.06946 [quant-ph]} \BibitemShut
  {NoStop}%
\bibitem [{\citenamefont {Bultink}\ \emph {et~al.}(2020)\citenamefont
  {Bultink}, \citenamefont {O{\textquoteright}Brien}, \citenamefont {Vollmer},
  \citenamefont {Muthusubramanian}, \citenamefont {Beekman}, \citenamefont
  {Rol}, \citenamefont {Fu}, \citenamefont {Tarasinski}, \citenamefont
  {Ostroukh}, \citenamefont {Varbanov}, \citenamefont {Bruno},\ and\
  \citenamefont {DiCarlo}}]{Bultink2020}%
  \BibitemOpen
  \bibfield  {author} {\bibinfo {author} {\bibfnamefont {C.~C.}\ \bibnamefont
  {Bultink}}, \bibinfo {author} {\bibfnamefont {T.~E.}\ \bibnamefont
  {O{\textquoteright}Brien}}, \bibinfo {author} {\bibfnamefont
  {R.}~\bibnamefont {Vollmer}}, \bibinfo {author} {\bibfnamefont
  {N.}~\bibnamefont {Muthusubramanian}}, \bibinfo {author} {\bibfnamefont
  {M.~W.}\ \bibnamefont {Beekman}}, \bibinfo {author} {\bibfnamefont {M.~A.}\
  \bibnamefont {Rol}}, \bibinfo {author} {\bibfnamefont {X.}~\bibnamefont
  {Fu}}, \bibinfo {author} {\bibfnamefont {B.}~\bibnamefont {Tarasinski}},
  \bibinfo {author} {\bibfnamefont {V.}~\bibnamefont {Ostroukh}}, \bibinfo
  {author} {\bibfnamefont {B.}~\bibnamefont {Varbanov}}, \bibinfo {author}
  {\bibfnamefont {A.}~\bibnamefont {Bruno}}, \ and\ \bibinfo {author}
  {\bibfnamefont {L.}~\bibnamefont {DiCarlo}},\ }\bibfield  {title} {\enquote
  {\bibinfo {title} {Protecting quantum entanglement from leakage and qubit
  errors via repetitive parity measurements},}\ }\href {\doibase
  10.1126/sciadv.aay3050} {\bibfield  {journal} {\bibinfo  {journal} {Science
  Advances}\ }\textbf {\bibinfo {volume} {6}} (\bibinfo {year} {2020}),\
  10.1126/sciadv.aay3050}\BibitemShut {NoStop}%
\bibitem [{\citenamefont {Bialczak}\ \emph {et~al.}(2010)\citenamefont
  {Bialczak}, \citenamefont {Ansmann}, \citenamefont {Hofheinz}, \citenamefont
  {Lucero}, \citenamefont {Neeley}, \citenamefont {O'Connell}, \citenamefont
  {Sank}, \citenamefont {Wang}, \citenamefont {Wenner}, \citenamefont
  {Steffen}, \citenamefont {Cleland},\ and\ \citenamefont
  {Martinis}}]{Bialczak2010}%
  \BibitemOpen
  \bibfield  {author} {\bibinfo {author} {\bibfnamefont {R.~C.}\ \bibnamefont
  {Bialczak}}, \bibinfo {author} {\bibfnamefont {M.}~\bibnamefont {Ansmann}},
  \bibinfo {author} {\bibfnamefont {M.}~\bibnamefont {Hofheinz}}, \bibinfo
  {author} {\bibfnamefont {E.}~\bibnamefont {Lucero}}, \bibinfo {author}
  {\bibfnamefont {M.}~\bibnamefont {Neeley}}, \bibinfo {author} {\bibfnamefont
  {A.~D.}\ \bibnamefont {O'Connell}}, \bibinfo {author} {\bibfnamefont
  {D.}~\bibnamefont {Sank}}, \bibinfo {author} {\bibfnamefont {H.}~\bibnamefont
  {Wang}}, \bibinfo {author} {\bibfnamefont {J.}~\bibnamefont {Wenner}},
  \bibinfo {author} {\bibfnamefont {M.}~\bibnamefont {Steffen}}, \bibinfo
  {author} {\bibfnamefont {A.~N.}\ \bibnamefont {Cleland}}, \ and\ \bibinfo
  {author} {\bibfnamefont {J.~M.}\ \bibnamefont {Martinis}},\ }\bibfield
  {title} {\enquote {\bibinfo {title} {Quantum process tomography of a
  universal entangling gate implemented with josephson phase qubits},}\ }\href
  {\doibase 10.1038/nphys1639} {\bibfield  {journal} {\bibinfo  {journal} {Nat.
  Phys.}\ }\textbf {\bibinfo {volume} {6}},\ \bibinfo {pages} {409--413}
  (\bibinfo {year} {2010})}\BibitemShut {NoStop}%
\bibitem [{\citenamefont {Dewes}\ \emph {et~al.}(2012)\citenamefont {Dewes},
  \citenamefont {Ong}, \citenamefont {Schmitt}, \citenamefont {Lauro},
  \citenamefont {Boulant}, \citenamefont {Bertet}, \citenamefont {Vion},\ and\
  \citenamefont {Esteve}}]{Dewes2012}%
  \BibitemOpen
  \bibfield  {author} {\bibinfo {author} {\bibfnamefont {A.}~\bibnamefont
  {Dewes}}, \bibinfo {author} {\bibfnamefont {F.~R.}\ \bibnamefont {Ong}},
  \bibinfo {author} {\bibfnamefont {V.}~\bibnamefont {Schmitt}}, \bibinfo
  {author} {\bibfnamefont {R.}~\bibnamefont {Lauro}}, \bibinfo {author}
  {\bibfnamefont {N.}~\bibnamefont {Boulant}}, \bibinfo {author} {\bibfnamefont
  {P.}~\bibnamefont {Bertet}}, \bibinfo {author} {\bibfnamefont
  {D.}~\bibnamefont {Vion}}, \ and\ \bibinfo {author} {\bibfnamefont
  {D.}~\bibnamefont {Esteve}},\ }\bibfield  {title} {\enquote {\bibinfo {title}
  {Characterization of a two-transmon processor with individual single-shot
  qubit readout},}\ }\href {\doibase 10.1103/PhysRevLett.108.057002} {\bibfield
   {journal} {\bibinfo  {journal} {Phys. Rev. Lett.}\ }\textbf {\bibinfo
  {volume} {108}},\ \bibinfo {pages} {057002} (\bibinfo {year}
  {2012})}\BibitemShut {NoStop}%
\bibitem [{\citenamefont {Strauch}\ \emph {et~al.}(2003)\citenamefont
  {Strauch}, \citenamefont {Johnson}, \citenamefont {Dragt}, \citenamefont
  {Lobb}, \citenamefont {Anderson},\ and\ \citenamefont
  {Wellstood}}]{Strauch2003}%
  \BibitemOpen
  \bibfield  {author} {\bibinfo {author} {\bibfnamefont {Frederick~W.}\
  \bibnamefont {Strauch}}, \bibinfo {author} {\bibfnamefont {Philip~R.}\
  \bibnamefont {Johnson}}, \bibinfo {author} {\bibfnamefont {Alex~J.}\
  \bibnamefont {Dragt}}, \bibinfo {author} {\bibfnamefont {C.~J.}\ \bibnamefont
  {Lobb}}, \bibinfo {author} {\bibfnamefont {J.~R.}\ \bibnamefont {Anderson}},
  \ and\ \bibinfo {author} {\bibfnamefont {F.~C.}\ \bibnamefont {Wellstood}},\
  }\bibfield  {title} {\enquote {\bibinfo {title} {Quantum logic gates for
  coupled superconducting phase qubits},}\ }\href {\doibase
  10.1103/PhysRevLett.91.167005} {\bibfield  {journal} {\bibinfo  {journal}
  {Phys. Rev. Lett.}\ }\textbf {\bibinfo {volume} {91}},\ \bibinfo {pages}
  {167005} (\bibinfo {year} {2003})}\BibitemShut {NoStop}%
\bibitem [{\citenamefont {DiCarlo}\ \emph {et~al.}(2010)\citenamefont
  {DiCarlo}, \citenamefont {Reed}, \citenamefont {Sun}, \citenamefont
  {Johnson}, \citenamefont {Chow}, \citenamefont {Gambetta}, \citenamefont
  {Frunzio}, \citenamefont {Girvin}, \citenamefont {Devoret},\ and\
  \citenamefont {Schoelkopf}}]{DiCarlo2010}%
  \BibitemOpen
  \bibfield  {author} {\bibinfo {author} {\bibfnamefont {L.}~\bibnamefont
  {DiCarlo}}, \bibinfo {author} {\bibfnamefont {M.~D.}\ \bibnamefont {Reed}},
  \bibinfo {author} {\bibfnamefont {L.}~\bibnamefont {Sun}}, \bibinfo {author}
  {\bibfnamefont {B.~R.}\ \bibnamefont {Johnson}}, \bibinfo {author}
  {\bibfnamefont {J.~M.}\ \bibnamefont {Chow}}, \bibinfo {author}
  {\bibfnamefont {J.~M.}\ \bibnamefont {Gambetta}}, \bibinfo {author}
  {\bibfnamefont {L.}~\bibnamefont {Frunzio}}, \bibinfo {author} {\bibfnamefont
  {S.~M.}\ \bibnamefont {Girvin}}, \bibinfo {author} {\bibfnamefont {M.~H.}\
  \bibnamefont {Devoret}}, \ and\ \bibinfo {author} {\bibfnamefont {R.~J.}\
  \bibnamefont {Schoelkopf}},\ }\bibfield  {title} {\enquote {\bibinfo {title}
  {Preparation and measurement of three-qubit entanglement in a superconducting
  circuit},}\ }\href {\doibase doi:10.1038/nature09416} {\bibfield  {journal}
  {\bibinfo  {journal} {Nature}\ }\textbf {\bibinfo {volume} {467}},\ \bibinfo
  {pages} {574--578} (\bibinfo {year} {2010})}\BibitemShut {NoStop}%
\bibitem [{\citenamefont {Andersen}\ \emph
  {et~al.}(2019{\natexlab{b}})\citenamefont {Andersen}, \citenamefont {Remm},
  \citenamefont {Lazar}, \citenamefont {Krinner}, \citenamefont {Lacroix},
  \citenamefont {Norris}, \citenamefont {Gabureac}, \citenamefont {Eichler},\
  and\ \citenamefont {Wallraff}}]{Andersen2019b}%
  \BibitemOpen
  \bibfield  {author} {\bibinfo {author} {\bibfnamefont {Christian~Kraglund}\
  \bibnamefont {Andersen}}, \bibinfo {author} {\bibfnamefont {Ants}\
  \bibnamefont {Remm}}, \bibinfo {author} {\bibfnamefont {Stefania}\
  \bibnamefont {Lazar}}, \bibinfo {author} {\bibfnamefont {Sebastian}\
  \bibnamefont {Krinner}}, \bibinfo {author} {\bibfnamefont {Nathan}\
  \bibnamefont {Lacroix}}, \bibinfo {author} {\bibfnamefont {Graham~J.}\
  \bibnamefont {Norris}}, \bibinfo {author} {\bibfnamefont {Mihai}\
  \bibnamefont {Gabureac}}, \bibinfo {author} {\bibfnamefont {Christopher}\
  \bibnamefont {Eichler}}, \ and\ \bibinfo {author} {\bibfnamefont {Andreas}\
  \bibnamefont {Wallraff}},\ }\bibfield  {title} {\enquote {\bibinfo {title}
  {Repeated quantum error detection in a surface code},}\ }\href
  {https://arxiv.org/abs/1912.09410} {\bibfield  {journal} {\bibinfo  {journal}
  {arXiv:1912.09410}\ } (\bibinfo {year} {2019}{\natexlab{b}})}\BibitemShut
  {NoStop}%
\bibitem [{\citenamefont {Koch}\ \emph {et~al.}(2007)\citenamefont {Koch},
  \citenamefont {Yu}, \citenamefont {Gambetta}, \citenamefont {Houck},
  \citenamefont {Schuster}, \citenamefont {Majer}, \citenamefont {Blais},
  \citenamefont {Devoret}, \citenamefont {Girvin},\ and\ \citenamefont
  {Schoelkopf}}]{Koch2007}%
  \BibitemOpen
  \bibfield  {author} {\bibinfo {author} {\bibfnamefont {J.}~\bibnamefont
  {Koch}}, \bibinfo {author} {\bibfnamefont {T.~M.}\ \bibnamefont {Yu}},
  \bibinfo {author} {\bibfnamefont {J.}~\bibnamefont {Gambetta}}, \bibinfo
  {author} {\bibfnamefont {A.~A.}\ \bibnamefont {Houck}}, \bibinfo {author}
  {\bibfnamefont {D.~I.}\ \bibnamefont {Schuster}}, \bibinfo {author}
  {\bibfnamefont {J.}~\bibnamefont {Majer}}, \bibinfo {author} {\bibfnamefont
  {A.}~\bibnamefont {Blais}}, \bibinfo {author} {\bibfnamefont {M.~H.}\
  \bibnamefont {Devoret}}, \bibinfo {author} {\bibfnamefont {S.~M.}\
  \bibnamefont {Girvin}}, \ and\ \bibinfo {author} {\bibfnamefont {R.~J.}\
  \bibnamefont {Schoelkopf}},\ }\bibfield  {title} {\enquote {\bibinfo {title}
  {Charge-insensitive qubit design derived from the {Cooper} pair box},}\
  }\href {\doibase 10.1103/PhysRevA.76.042319} {\bibfield  {journal} {\bibinfo
  {journal} {Phys. Rev. A}\ }\textbf {\bibinfo {volume} {76}},\ \bibinfo {eid}
  {042319} (\bibinfo {year} {2007})}\BibitemShut {NoStop}%
\bibitem [{\citenamefont {Terhal}(2015)}]{Terhal2015n}%
  \BibitemOpen
  \bibfield  {author} {\bibinfo {author} {\bibfnamefont {Barbara~M.}\
  \bibnamefont {Terhal}},\ }\bibfield  {title} {\enquote {\bibinfo {title}
  {Quantum error correction for quantum memories},}\ }\href {\doibase
  10.1103/RevModPhys.87.307} {\bibfield  {journal} {\bibinfo  {journal} {Rev.
  Mod. Phys.}\ }\textbf {\bibinfo {volume} {87}},\ \bibinfo {pages} {307--346}
  (\bibinfo {year} {2015})}\BibitemShut {NoStop}%
\bibitem [{\citenamefont {Reiher}\ \emph {et~al.}(2017)\citenamefont {Reiher},
  \citenamefont {Wiebe}, \citenamefont {Svore}, \citenamefont {Wecker},\ and\
  \citenamefont {Troyer}}]{Reiher2017}%
  \BibitemOpen
  \bibfield  {author} {\bibinfo {author} {\bibfnamefont {Markus}\ \bibnamefont
  {Reiher}}, \bibinfo {author} {\bibfnamefont {Nathan}\ \bibnamefont {Wiebe}},
  \bibinfo {author} {\bibfnamefont {Krysta~M.}\ \bibnamefont {Svore}}, \bibinfo
  {author} {\bibfnamefont {Dave}\ \bibnamefont {Wecker}}, \ and\ \bibinfo
  {author} {\bibfnamefont {Matthias}\ \bibnamefont {Troyer}},\ }\bibfield
  {title} {\enquote {\bibinfo {title} {Elucidating reaction mechanisms on
  quantum computers},}\ }\href {\doibase 10.1073/pnas.1619152114} {\bibfield
  {journal} {\bibinfo  {journal} {Proceedings of the National Academy of
  Sciences}\ }\textbf {\bibinfo {volume} {114}},\ \bibinfo {pages} {7555--7560}
  (\bibinfo {year} {2017})}\BibitemShut {NoStop}%
\bibitem [{\citenamefont {Nielsen}\ and\ \citenamefont
  {Chuang}(2011)}]{Nielsen2011}%
  \BibitemOpen
  \bibfield  {author} {\bibinfo {author} {\bibfnamefont {M.~A.}\ \bibnamefont
  {Nielsen}}\ and\ \bibinfo {author} {\bibfnamefont {I.~L.}\ \bibnamefont
  {Chuang}},\ }\href@noop {} {\emph {\bibinfo {title} {Quantum Computation and
  Quantum Information}}},\ \bibinfo {edition} {10th}\ ed.\ (\bibinfo
  {publisher} {Cambridge University Press},\ \bibinfo {address} {New York, NY,
  USA},\ \bibinfo {year} {2011})\BibitemShut {NoStop}%
\end{thebibliography}%
\end{document}